\documentclass{article}

\usepackage[backend=biber,sorting=none,style=trad-abbrv,maxbibnames=1]{biblatex}

\usepackage[subpreambles=true]{standalone}
\usepackage{hyperref}
\usepackage{ragged2e}

\usepackage{theme}

\usepackage{cleveref}
\crefname{equation}{Eq.}{Eqs.}
\crefname{figure}{Fig.}{Figs.}
\crefname{section}{Sec.}{Secs.}
\Crefname{section}{Sec.}{Secs.}
\crefname{appendix}{App.}{Apps.}
\Crefname{appendix}{App.}{Apps.}
\crefname{table}{Table}{Tables}

\AtEveryBibitem{
    \clearfield{pagetotal}
    \clearfield{isbn}
    \clearfield{issn}
    \clearfield{eprint}
    \clearfield{eprinttype}
}

\begin{document}

\maketitle

\begin{abstract}
    In this paper, we present various well-studied examples of quantum systems, and discuss how they might be analysed algebraically. We analyse the harmonic oscillator, hydrogenic atom, optical components, and particle spin using multiple algebraic methods, and highlight the central importance of algebraic relationships between operators in determining the physical character of the system. In particular, we show: the ubiquity of the Lie algebra $\sotwoone\cong\suoneone$ as the ``spectrum generating algebra''; the role of $\sutwo$ and $\suoneone$ in determining the action of optical components; that an elementary algebraic approach to particle spin offers novel insights into its nature versus the usual approaches; and how understanding the full (often hidden) symmetry of a problem enables it to be solved immediately without the need to solve differential equations. We hope this work will serve as a gentle introduction to algebraic methods in quantum mechanics for the interested reader --- we therefore signpost active areas of research within the community.
\end{abstract}

\section{Introduction}

\subsection{Aim}

The aim of this paper is to highlight some deep links between quantum mechanics and its algebraic structure which is otherwise obscured by the formalism of the \schrodinger\space picture. To do this, we will carefully examine familiar examples of quantum systems from an algebraic perspective. In many cases, we will demonstrate their direct solution from this structure without needing to solve a differential equation nor find explicit states for the system. In so doing, we will show how an understanding of algebraic structure yields new insights into already well-understood quantum phenomena, offering a path to further discovery beyond the examples we discuss.

\subsection{Algebras and Quantum Mechanics}

Quantum mechanics and quantum field theory are widely considered to be the most successful physical theories humankind has produced to date. In a first course, and often beyond, the tradition within physics sees the student of quantum mechanics taught the \schrodinger\space picture with position dependent wavefunctions acted upon by a variety of differential operators. They will be taught to solve the partial differential equations of the \schrodinger\space equation for a number of classic systems, especially the quantum harmonic oscillator and the hydrogenic atom. Over the course of this instruction, the student will observe the emergence of discrete energy eigenstates from the technical complexities of the formalism, and marvel that such discreteness can emerge from a manifestly continuum theory.

Meanwhile, the student will be introduced to the canonical commutation relations encoding the relationship between position and momentum operators,
\begin{equation}\label{eqn:heisenberg-algebra}
    \left[\hat{x}_{j},\hat{x}_{k}\right]=\left[\hat{p}_{j},\hat{p}_{k}\right]=0,\quad\left[\hat{x}_{j},\hat{p}_{k}\right]=\unitimaginary\hbar\delta_{jk}.
\end{equation}
They will learn that there is a connection between this relationship and the Heisenberg uncertainty principle, and how to represent each operator with respect to the other's eigenbasis. When solving the quantum harmonic oscillator, they will be taught to form a strange linear combination of position and momentum operators,
\begin{equation}
    \definedas{\hat{a}_{j}}{\sqrt{\frac{m\omega}{2\hbar}}\hat{q}_{j}+\frac{\unitimaginary}{\sqrt{2m\omega\hbar}}\hat{p}_{j}},
\end{equation}
which, together with its Hermitian conjugate allows them to rapidly solve the problem. They will learn that such a ``ladder operator'' method may be applied to the hydrogen atom, and is essential in quantum field theory to handle creation and annihilation of particles.

Later, they will learn about spin-$\half$ systems and that the operators required for its description $\set{\hat{S}_{1},\hat{S}_{2},\hat{S}_{3}}$ are familiar finite-dimensional matrices. They will be taught how these operators combine,
\begin{equation}\label{eqn:spin-half-relation}
    \hat{S}_{a}\hat{S}_{b}=\frac{\hbar^{2}}{4}\delta_{ab}I_{2}+\half[\unitimaginary\hbar]\sothreesum{c}\sothreestructureconstants{a}{b}{c}\hat{S}_{c},
\end{equation}
with $I_{2}$ the identity matrix, and may learn the form of operators with higher spin than $\half$, noting that \Cref{eqn:spin-half-relation} no longer holds.

What the student may not realise, with these details of differential operators and matrices obscuring their vision, is the central importance of \Cref{eqn:heisenberg-algebra} and \Cref{eqn:spin-half-relation}, each of which constitute \textit{algebraic} relationships, in giving \textit{form} to the systems they are studying. To understand what we mean, and to motivate the study of quantum mechanics by algebraic means more generally, let us consider the underlying physical concepts of the examples we have mentioned in more detail.

\subsection{Position and Momentum}

In a 1925 paper by Dirac~\cite{dirac_fundamental_1925}, he used the new ``quantum mechanics'' developed by Heisenberg~\cite{heisenberg_uber_1925} (known today as ``matrix mechanics'') to derive the canonical commutation relations between position and momentum \Cref{eqn:heisenberg-algebra}. Sadly for him, they had been derived using matrix mechanics by Born and Jordan~\cite{born_zur_1925} a little over a month earlier, and these are today known as the ``Heisenberg algebra''. Unlike the common view today, Dirac considered these position and momentum objects not as operators but as ``q-numbers'': a generalisation of ordinary ``c-numbers'' obeying all the usual laws of arithmetic except commutativity. In his 1926 work on the hydrogen spectrum~\cite{dirac_quantum_1926}, Dirac highlights that this difference is what enables atomic spectra to be successfully derived in quantum mechanics but not in classical mechanics.

Dirac's insight anticipated later work which showed just how important these relations are in making quantum mechanics quantum at all. In the theory of `Koopman-von Neumann classical physics''~\cite{piasecki_introduction_2021}, classical mechanics is mathematically formulated using what is usually considered quantum language: on a complex Hilbert space with complex-valued wavefunctions, probability amplitudes, non-commutative operators, and physical observables corresponding to Hermitian operators. That this does not reproduce quantum physics, despite the identical mathematical setting, can be traced to one assumption: in Koopman-von Neumann classical physics, the position and momentum operators commute~\cite{bondar_operational_2012}:
\begin{equation}\label{eqn:heisenberg-algebraKOOPMANVONNEUMANN}
    \left[\hat{x}_{j},\hat{x}_{k}\right]=\left[\hat{p}_{j},\hat{p}_{k}\right]=\left[\hat{x}_{j},\hat{p}_{k}\right]=0.
\end{equation}
Remarkably, one can only make the \textit{classical} theory work if new operators which \textit{don't} commute with position and momentum are introduced,
\begin{equation}
    \bigl[\hat{x}_{j},\hat{\lambda}_{x_{k}}\bigr]=\bigl[\hat{p}_{j},\hat{\lambda}_{p_{k}}\bigr]=\unitimaginary\delta_{jk}.
\end{equation}

This model reveals that non-commutativity in itself is not enough to develop quantum mechanics --- indeed it is essential to both classical and quantum mechanics! Instead, it is the non-commutative relationship between position and momentum specifically, as encoded in \Cref{eqn:heisenberg-algebra}, which yields quantum mechanics. That is to say, it is the algebraic structure between the observables in the theory which determines its predictions, and these stand independently of the formalism used to realise them. Particularly within \schrodinger's wave mechanics, where all focus is on the wavefunction, this critical structure is obscured.

\subsection{Spin Angular Momentum}

Spin is understood to be an intrinsic form of angular momentum possessed by physical systems, which exists in amounts of $\definedas{s}{\half[k]}\hbar$ with $k\in\naturals$, which then allows the system $2s+1$ possible eigenstates of projection along a given spatial axis. Its signature had been seen in atomic spectra, in particular through the anomalous Zeeman effect, for decades before Pauli's initial explanatory model of its phenomenology through an inherently quantum ``two-valuedness''~\cite{pauli_uber_1923}. By 1927, the nature of this ``two-valuedness'' was settled as a form of angular momentum, after which Pauli introduced his eponymous matrices for the description of electron spin~\cite{pauli_zur_1927},
\begin{equation}
    \definedas{\sigma_{x}}{\paulix},\quad\definedas{\sigma_{y}}{\pauliy},\quad\definedas{\sigma_{z}}{\pauliz}.
\end{equation}

Working instead with the spin generators $\definedas{\hat{S}_{a}}{\half[\hbar]\sigma_{a}}$, one may confirm that \Cref{eqn:spin-half-relation} is correct. With some careful consideration, one may conclude that \Cref{eqn:spin-half-relation} also provides enough information to recover the matrix form of the $\set{\hat{S}_{a}}$ up to labelling. This equation may be decomposed into two parts: the first are commutation relations between the $\set{\hat{S}_{a}}$,
\begin{equation}\label{eqn:spin-commutation-relations}
    \bigl[\hat{S}_{a},\hat{S}_{b}\bigr]=\unitimaginary\hbar\sothreesum{c}\sothreestructureconstants{a}{b}{c}\hat{S}_{c},
\end{equation}
which reflect the non-commutativity of classical rotations; the second is an example of the defining equation of a ``Clifford algebra''~\cite{doran_geometric_2003},
\begin{equation}\label{eqn:spin-half-clifford}
    \hat{S}_{a}\hat{S}_{b}+\hat{S}_{b}\hat{S}_{a}=\frac{\hbar^{2}}{4}\delta_{ab}I_{2},
\end{equation}
and contains the spectral information of the spin generators. We see that both algebraic relations are needed to fully characterise the spin degrees of freedom.

Matrix descriptions of higher-spin systems are known \cite{binney_physics_2013} from the representation theory of the Lie group $\specialunitarygroup{2}$, and are widely used. For such systems \Cref{eqn:spin-commutation-relations} is always satisfied; what is less well known is that \Cref{eqn:spin-half-clifford} has higher-order analogues in these systems. For example, for the spin-$1$ system with matrix generators,
\begin{equation}
    \hat{S}_{x}=\spinxone,\quad\hat{S}_{y}=\spinyone,\quad\hat{S}_{z}=\spinzone,
\end{equation}
we find that they satisfy the defining equation of the ``Duffin-Kemmer-Petiau algebra'' \cite{helmstetter_about_2010,micali_meson_2008},
\begin{equation}\label{eqn:spin-one-dkp}
    \hat{S}_{a}\hat{S}_{b}\hat{S}_{c}+\hat{S}_{c}\hat{S}_{b}\hat{S}_{a}=\hbar^{2}\delta_{bc}\hat{S}_{a}+\hbar^{2}\delta_{ab}\hat{S}_{c}.
\end{equation}
Like \Cref{eqn:spin-half-clifford}, \Cref{eqn:spin-one-dkp} contains the spectral information of the spin generators, and together with \Cref{eqn:spin-commutation-relations} is sufficient to reconstruct the matrix representation up to labelling. Working with these algebraic relationships (instead of matrices) has the advantage that we may avoid the choice of a spinor basis in the problem. From this perspective, we also note that both \Cref{eqn:spin-half-clifford,eqn:spin-one-dkp} are \textit{real}, and each term has the same power of spin generators modulo 2. This means that we may move to a generator basis in which all of \Cref{eqn:spin-commutation-relations,eqn:spin-half-clifford,eqn:spin-one-dkp} are also \textit{real}. Despite Pauli's assertion that the ``two-valuedness'' of spin is inherently quantum in nature, this observation offers the possibility of connecting spin more concretely with classical rotational symmetry via a real algebraic description.

\subsection{Creation and Annihilation Operators}

The history of creation and annihilation operators in quantum mechanics is nearly as old as quantum mechanics itself~\cite{rushka_completely_2020}. Using them, one may find a family of eigenstates of a collection of operators given some other eigenstate. When the operator in question is a Hamiltonian with discrete spectrum, one finds the energy spectrum of the problem. If instead one considers the $\hat{S}_{z}$ operator, one finds all possible projections of the spin along the $z$-directions. Indeed, this method is core to the representation theory of complex semisimple Lie algebras~\cite{fulton_representation_1991}, the results of which are widely used in particle physics.

Moreover, it is through creation and annihilation operators that one may most conveniently encode statistical information about a particle species. Specifically, those particles obeying Bose-Einstein statistics satisfy commutation relations,
\begin{equation}
    \bigl[\hat{a}_{j},\hat{a}_{k}\bigr]=\bigl[\hat{a}^{\dagger}_{j},\hat{a}^{\dagger}_{k}\bigr]=0,\quad\bigl[\hat{a}_{j},\hat{a}^{\dagger}_{k}\bigr]=\delta_{jk},
\end{equation}
while those satisfying Fermi-Dirac statistics satisfy anticommutation relations,
\begin{equation}
    \bigl\{\hat{a}_{j},\hat{a}_{k}\bigr\}=\bigl\{\hat{a}^{\dagger}_{j},\hat{a}^{\dagger}_{k}\bigr\}=0,\quad\bigl\{\hat{a}_{j},\hat{a}^{\dagger}_{k}\bigr\}=\delta_{jk}.
\end{equation}
A difference in sign in these algebraic relations is enough to either allow particles to occupy the same quantum states or not. Together with the spin-statistics theorem, first proved by Fierz \cite{fierz_uber_1939} and developed by Pauli \cite{pauli_connection_1940}, this sign difference is responsible for the Pauli exclusion principle and ultimately the properties of the material world. Perhaps more tangibly, many uniquely quantum effects that present in our experiments such as the ``Hong-Ou-Mandel'' effect \cite{bouchard_two-photon_2020,Hance_2025}, originate from particle statistics and can be understood most readily through the creation and annihilation operators.

\subsection{Outline}

In \Cref{sec:harmonic-oscillator} we will begin our discourse with the humble harmonic oscillator, considering it from a number of different algebraic perspectives, all offering methods of direct solution. We will also find an example of hidden $\sotwoone$ and $\suthree$ structure mathematically relating its energy eigenstates. Next, in \Cref{sec:hydrogen} we consider the hydrogenic atom, discussing Pauli's original algebraic derivation within matrix mechanics, revealing a hidden $\specialorthogonalgroup{4}$ symmetry and identifying the same ``spectrum generating'' structure as in the harmonic oscillator problem. In \Cref{sec:optics}, we then consider the ways in which the behaviour of ubiquitous optical components are determined by $\sutwoc$ and $\suoneonec$ algebras, and how spin statistics via annihilation and creation operators give rise to uniquely quantum optical effects. We present our final example in \Cref{sec:spin-algebras} where we consider the problem of particle spin using three different but related algebraic methods. Confirming the existence of real algebraic descriptions underlying such systems, we discuss how these algebras offer greater insight into the physical properties and elementary nature of such systems. We signpost to some active areas of algebraic research in quantum mechanics and provide concluding remarks in \Cref{sec:discussion-conclusion}.

Note: we have consistently used hats in this introduction when discussing familiar operators in the quantum setting; throughout the rest of the paper these shall be dropped, as when considering algebraic structure these objects do not necessarily ``operate'' on anything!

\section{Harmonic Oscillator}\label{sec:harmonic-oscillator}

Where else to begin if not with the humble harmonic oscillator? A mainstay of theoretical physics, harmonic oscillators are essential models for describing physical phenomena and the perfect test-bed for demonstrating perhaps unfamiliar mathematical paradigms, such as algebraic approaches to eigenspectra. We will derive the eigenspectrum of the quantum harmonic oscillator thrice: once using the familiar ladder operators, once utilising the structure of the Lie algebra $\suoneonec$, and a final time using the hidden $\suthreec$ symmetry possessed by the three-dimensional problem.

\subsection{Ladder Operators}

A discussion of algebraic methods in quantum mechanics that does not feature ladder operators would be incomplete at best (and wilfully ignorant at worst). When a system of ladder operators are known for a physical system, they can greatly simplify many calculations, allow one to find explicit forms for eigenstates with ease, and are indispensable tools in quantum field theory. In fact, the treatment of the harmonic oscillator by ladder operators is nearly as old as quantum mechanics itself through the work of Born, Jordan, and Dirac~\cite{rushka_completely_2020}. The presentation we give here is well-known but consult Messiah~\cite{messiah_quantum_1970} for more details.

To begin, first recall the familiar Hamiltonian in Cartesian coordinates for an isotropic simple harmonic oscillator in $D$-dimensions,
\begin{equation}\label{eqn:harmonic-oscillator-hamiltonian}
    H=\sum_{j=1}^{D}\left(\frac{1}{2m}p_{j}^{2}+\frac{1}{2}m\omega^{2}q_{j}^{2}\right).
\end{equation}
Let us now define a new set of operators,
\begin{equation}
    \definedas{a_{j}}{\sqrt{\frac{m\omega}{2\hbar}}q_{j}+\frac{\unitimaginary}{\sqrt{2m\omega\hbar}}p_{j}},
\end{equation}
which can easily be shown to satisfy, $\forall j,k\in\set{1,\dots,D}$
\begin{equation}
\begin{gathered}
    \commutator{a_{j}}{a_{k}}=\commutator{a^{\dagger}_{j}}{a^{\dagger}_{k}}=0\\
    \commutator{a_{j}}{a^{\dagger}_{k}}=\delta_{jk}.
\end{gathered}
\end{equation}
Writing $q_{j}$ and $p_{j}$ in terms of $a_{j}$ and $a^{\dagger}_{j}$, we find,
\begin{equation}
    \definedas{H=\half[\hbar\omega]\sum_{j=1}^{D}\left(a_{j}a^{\dagger}_{j}+a^{\dagger}_{j}a_{j}\right)=\hbar\omega\sum_{j=1}^{D}\left(a^{\dagger}_{j}a_{j}+\half\right)}{\hbar\omega\sum_{j=1}^{D}H_{j}}.
\end{equation}
Each term in the sum commutes with all others and thus with the Hamiltonian; this means we may decompose an energy eigenstate into mutual eigenstates of the $\set{\definedas{N_{j}}{a^{\dagger}_{j}a_{j}}}$ operators.

To pin down the eigenspectrum, notice that,
\begin{equation}\label{eqn:ladder-operator-number-operator-relations}
\begin{gathered}
    N_{j}a_{k}=a_{k}(N_{j}-\delta_{jk})\\
    N_{j}a^{\dagger}_{k}=a^{\dagger}_{k}(N_{j}+\delta_{jk}),
\end{gathered}
\end{equation}
and so $a_{j}$ and $a^{\dagger}_{j}$ lower and raise respectfully the eigenvalue of an eigenstate of $N_{j}$ by one. At this point in the derivation, it is customary to appeal to the Hilbert space structure of the states of the system. Denoting,
\begin{equation}
    N_{j}\ket{\nu}=\nu\ket{\nu},
\end{equation}
consider the Hilbert space norm,
\begin{equation}
    \definedas{\norm{\,a_{j}\ket{\nu}}^{2}}{\bra{\nu}a^{\dagger}_{j}a_{j}\ket{\nu}}=\bra{\nu}N_{j}\ket{\nu}=\nu\braket{\nu\vert\nu}\geq0.
\end{equation}
Since each $N_{j}$ is Hermitian, its eigenvalues are necessarily real, and so from the above we must have $\nu\in\reals_{\geq0}$. If $\nu\notin\naturals$, then by \Cref{eqn:ladder-operator-number-operator-relations} there must exist $M\in\pintegers$ such that $N_{j}a^{M}_{j}\ket{\nu}=(\nu-M)a^{M}_{j}\ket{\nu}$ with $\nu-M<0$. Therefore, $a^{M}_{j}\ket{\nu}=0$, and $(\nu-M+1)a^{M-1}_{j}\ket{\nu}=N_{j}a^{M-1}_{j}\ket{\nu}=a^{\dagger}_{j}a^{M}_{j}\ket{\nu}=0$; by induction we have $\forall K\in\pintegers, a^{K}_{j}\ket{\nu}=0$. Finally, this means $\nu\ket{\nu}=N_{j}\ket{\nu}=a^{\dagger}_{j}a_{j}\ket{\nu}=0$, and thus $\ket{\nu}=0$, contradicting our assumption that $\ket{\nu}$ was an eigenstate of $N_{j}$. This means $\nu\in\naturals$. It is worth noting that the authors have not seen the latter part of the proof of the spectrum of $N_{j}$ in any other treatments of the harmonic oscillator despite its essential role in ruling out non-natural eigenvalues. Now labelling the eigenvalues $\set{n_{j}\in\naturals}$ on a mutual eigenstate of $\set{N_{j}}$, we apply the Hamiltonian and immediately find the energy eigenspectrum,
\begin{equation}
    E=\hbar\omega\left(\half[D]+\sum_{j=1}^{D}n_{j}\right)=\hbar\omega\left(\half[D]+n\right)
\end{equation}
where we have introduced the usual quantum number $n\in\naturals$ for this system.

Ref.~\cite{jordan_quantum_1986} offers a variation on the final part of this derivation which is more algebraic in character with less focus on states once some operator properties are established. First, we easily verify, $\forall k\in\pintegers$,
\begin{equation}
    \commutator{(a_{j}^{\dagger})^{k}}{a_{j}}=-k(a_{j}^{\dagger})^{k-1}.
\end{equation}
Using this, we may prove by induction that,
\begin{equation}
    (a^{\dagger}_{j})^{k}(a_{j})^{k}=\prod_{m=1}^{k}\left(\frac{1}{\hbar\omega}H_{j}-\half[2m-1]\right),
\end{equation}
and that $(a^{\dagger}_{j})^{k}(a_{j})^{k}$ is a positive semidefinite operator for all $j\in\set{1,\dots,D}$, $k\in\pintegers$:
\begin{equation}
    \forall \psi,\quad\bra{\psi}(a^{\dagger}_{j})^{k}(a_{j})^{k}\ket{\psi}=\left((a_{j})^{k}\ket{\psi}\right)^{\dagger}(a_{j})^{k}\ket{\psi}\geq0.
\end{equation}
This entails $H_{j}$ is also positive semidefinite. Thus, taking expectation values with an eigenstate $\ket{E_{j}}$ of $H_{j}$ we must have,
\begin{equation}
    \prod_{m=1}^{k}\left(\frac{1}{\hbar\omega}E_{j}-\half[2m-1]\right)\geq0.
\end{equation}
As the second term in each factor of this product grows without bound with $k$, for this inequality to hold $\forall k\in\pintegers$ there must exist some least $k'\in\pintegers$ for each $E_{j}$ such that,
\begin{equation}
    \left(\frac{1}{\hbar\omega}E_{j}-\half[2k'-1]\right)=0,
\end{equation}
terminating the product before it can become negative. Thus, more conventionally writing $k'=n_{j}+1$ with $n_{j}\in\naturals$, we recover,
\begin{equation}
    E_{j}=\hbar\omega\left(n_{j}+\half\right).
\end{equation}

\subsection{The ``Spectrum Generating Algebra'' \texorpdfstring{$\suoneonec$}{su(1,1)C}}\label{sec:spectrum-generating-algebra}

Methods involving ladder operators are both ubiquitous, and in some ways are the most mysterious, used in quantum physics. Once known, we are able to render the discrete part of the spectrum of our operators of interest almost instantly, but it is not always clear from where these operators originated. In the case of the harmonic oscillator, the form of the potential enabled us to easily split the Hamiltonian into commuting parts, but this is not always possible. As non-self-adjoint operators, not corresponding to any physical observable themselves, their use is of great mathematical convenience but grants minimal physical insight into the system under study. However, another algebraic approach to deriving the spectrum of the harmonic oscillator is possible which directly uses self-adjoint operators associated with the system. Moreover, it enables us to derive the full spectrum of the Hamiltonian without needing to decompose it into commuting parts, at a relatively constant level of complexity.

To do this, we must consider the Lie algebra $\sotwoone\cong\suoneone$~\cite{iachello_lie_2015}; this algebra recurs frequently in physical problems and is a simple example of a ``spectrum generating algebra''~\cite{cordero_analysis_2005,levai_solvable_1994,dothan_finite-dimensional_1970}. We may define its complexification $\suoneonec$ on a basis $\set{\sotwoonegen[1],\sotwoonegen[2],\sotwoonegen[3]}$ with Lie product,
\begin{equation}\label{eqn:su-one-one-c}
    \commutator{\sotwoonegen[1]}{\sotwoonegen[2]}=-\unitimaginary\sotwoonegen[3],\quad
    \commutator{\sotwoonegen[2]}{\sotwoonegen[3]}=\unitimaginary\sotwoonegen[1],\quad
    \commutator{\sotwoonegen[3]}{\sotwoonegen[1]}=\unitimaginary\sotwoonegen[2].
\end{equation}
Like $\sutwoc$, this Lie algebra has one non-trivial Casimir element,
\begin{equation}
    \definedas{\sotwoonecasimir}{-\sotwoonegen[1]^{2}-\sotwoonegen[2]^{2}+\sotwoonegen[3]^{2}},
\end{equation}
and one may easily verify $\forall j\in\set{1,2,3}$,
\begin{equation}
    \commutator{\sotwoonecasimir}{\sotwoonegen[j]}=0.
\end{equation}
As with all low-dimensional algebraic structures, $\suoneonec$ is realised in myriad ways in a wide variety of larger algebraic structures. To get the most out of this structure, it is useful to find a realisation of it which is valid for arbitrary dimension $D$.

For the harmonic oscillator with Hamiltonian \eqref{eqn:harmonic-oscillator-hamiltonian}, we may easily verify that,
\begin{equation}\label{eqn:harmonic osciallator-realisation}
\begin{gathered}
    \sotwoonegen[1]=\frac{1}{2\hbar\omega}\sum_{j=1}^{D}\left(\frac{1}{2m}p_{j}^{2}-\frac{1}{2}m\omega^{2}q_{j}^{2}\right)\\
    \sotwoonegen[2]=\frac{1}{4\hbar}\sum_{j=1}^{D}\left(q_{j}p_{j}+p_{j}q_{j}\right)\\
    \sotwoonegen[3]=\frac{1}{2\hbar\omega}\sum_{j=1}^{D}\left(\frac{1}{2m}p_{j}^{2}+\frac{1}{2}m\omega^{2}q_{j}^{2}\right),
\end{gathered}
\end{equation}
satisfies the commutation relations \eqref{eqn:su-one-one-c}. It is worth noting that in this case that, up to a scale factor, $\sotwoonegen[1]$ is the Lagrangian of the system and $\sotwoonegen[3]$ is the Hamiltonian, but this need not be the case in general. The realisation of $\sotwoonegen[2]$ is also noteworthy: its classical counterpart is the canonical starting point for proof of the virial theorem~\cite{goldstein_classical_2002}, and the quantum version is sometimes referred to as the ``Berry-Keating'' operator~\cite{bagarello_berry-keating_2026}, for its central role in the conjecture of the same name, which seeks to connect the eigenvalues of this operator to the non-trivial zeroes along the $\realpart{z}=\half$ line of the Riemann Zeta function~\cite{wang_overviews_2020}. Unfortunately for us, the harmonic oscillator requires that we instead solve a $\sotwoonegen[3]$ eigenvalue problem.

It is clear from its construction that $\sotwoonegen[3]$ is a Hermitian and positive semi-definite operator with respect to the usual Hilbert space inner product, meaning its spectrum must be real and non-negative. Expressing the Casimir element $\sotwoonecasimir$ in our realisation \eqref{eqn:harmonic osciallator-realisation}, we find,
\begin{equation}
    \sotwoonecasimir=\frac{1}{4\hbar^{2}}L^{2}+\frac{1}{16}D(D-4),
\end{equation}
where $L^{2}$ is the Casimir element for the complexified $D$-dimensional Euclidean orbital angular momentum Lie algebra $\sodc$,
\begin{subequations}
\begin{gather}
    \definedas{L^{2}}{\half[\hbar^{2}]\smashoperator[r]{\sum_{j,k=1}^{D}}L_{jk}L_{jk}}\label{eqn:d-dim-orbital-cas}\\
    \definedas{L_{jk}}{x_{j}p_{k}-x_{k}p_{j}}\label{eqn:d-dim-orbital}
\end{gather}
\end{subequations}
which satisfies,
\begin{subequations}
\begin{gather}
    \commutator{L^{2}}{L_{jk}}=0\\
    \commutator{L_{jk}}{L_{lm}}=\unitimaginary\hbar\bigl(\delta_{jl}L_{km}+\delta_{km}L_{jl}-\delta_{jm}L_{kl}-\delta_{kl}L_{jm}\bigr).\label{eqn:d-dim-orb-comm}
\end{gather}
\end{subequations}
Since $\sotwoonecasimir$ commutes with $\sotwoonegen[3]$ so does $L^{2}$, and it is natural (due to the isotropy) to consider eigenstates of orbital angular momentum. In $D$-dimensions, such a mutual eigenstate satisfies~\cite{adams_algebraic_1994},
\begin{equation}
    L^{2}\ket{l,q}=l(l+D-2)\hbar^{2}\ket{l,q},\quad
    \sotwoonegen[3]\ket{l,q}=q\ket{l,q},
\end{equation}
and we have,
\begin{equation}
    \sotwoonecasimir\ket{l,q}=\frac{1}{16}(2l+D-4)(2l+D)\ket{l,q}.
\end{equation}
It is customary to write the value of $\sotwoonecasimir$ in the form $k(k+1)$; doing so, we find two possible values,
\begin{equation}
    k_{\pm}=\half\left(-1\pm\half(2l+D-2)\right).
\end{equation}
The unitary representations have been completely classified~\cite{biedenharn_unitary_1965,montgomery_noncompact_1974,barut_non-compact_1965}, and an exclusively non-negative $\sotwoonegen[3]$ spectrum can only occur when $k<0$, giving $\sotwoonegen[3]$ a discrete spectrum. For $D\geq2$ we must take $k_{-}$, and this spectrum is, $\forall\mu\in\naturals$,
\begin{equation}
    q=-k_{-}+\mu=\half l+\frac{1}{4}D+\mu.
\end{equation}
Therefore, since $\sotwoonegen[3]=\frac{1}{2\hbar\omega}H$, we find,
\begin{equation}
    E=\hbar\omega\left(n+\half D\right),
\end{equation}
where we have defined the quantum number $n=l+\mu\in\naturals$. As noted by Ref.~\cite{adams_algebraic_1994}, when $D=1$, the orbital angular momentum operators are identically zero, and both $k_{\pm}$ are valid, yielding,
\begin{equation}
    q_{\pm}=\frac{1}{2}\pm\frac{1}{4}+\mu,
\end{equation}
and so the energies,
\begin{equation}
    E_{\pm}=\hbar\omega\left(1\pm\half+2\mu\right).
\end{equation}
These correspond to odd and even quantum number solutions to the energy spectrum respectively, and can be combined. Thus, our treatment of the harmonic oscillator using $\suoneonec$ is in full agreement with the ladder operator method in all dimensions.

\subsection{The Fradkin Tensor and \texorpdfstring{$\sudc$}{su(d)C} Hidden Symmetry}

A common concern for physicists is determining which quantities are invariant under the evolution of the system. When such invariants exist, they offer a powerful method of solution, which yields valuable insights at the heart of the problem. The quantum harmonic oscillator is no exception. Indeed, we have already seen that a $D$-dimensional isotropic oscillator can be decomposed into commuting $1$-dimensional oscillators, offering one such insight. In that approach, these were the only Hermitian invariants. Another approach, that of the ``Fradkin tensor'', consists of an entire algebra of Hermitian invariants hiding a symmetry for the system in plain sight. Our approach will be similar to Hughes~\cite{hughes_harmonic_1973}, but will use a different basis to make the Casimir invariants much easier to handle.

If we were to break up the isotropic harmonic oscillator Hamiltonian into parts labelled by a pair of spatial indices and of the form,
\begin{equation}\label{eqn:fradkin}
    \definedas{\fradkin{jk}}{\frac{1}{2m}p_{j}p_{k}+\frac{1}{2}m\omega^{2}q_{j}q_{k}},
\end{equation}
for all $j,k\in\set{1,\dots,D}$, one would quickly discover that each of these operators commutes with the Hamiltonian, which could be described as,
\begin{equation}
    H=\sum_{j=1}^{D}\fradkin{jj}.
\end{equation}
Jauch and Hill~\cite{jauch_problem_1940} first identified this invariant in 1940 for the two-dimensional isotropic harmonic oscillator problem, which was later generalised to three-dimensions by Fradkin~\cite{fradkin_three-dimensional_1965} in 1965. Today, these tensors bear Fradkin's name, and one can easily show that each Fradkin tensor~\eqref{eqn:fradkin}, as well as the each of the $D$-dimensional orbital angular momentum operators~\eqref{eqn:d-dim-orbital}, are invariant for the $D$-dimensional isotropic problem. Together with \Cref{eqn:d-dim-orb-comm}, the $\set{L_{jk},\fradkin{mn}}$ form a Lie algebra,
\begin{equation}\label{eqn:orb-frad-comm-rels}
\begin{split}
    \commutator{L_{jk}}{L_{mn}}&=\unitimaginary\hbar\bigl(\delta_{jm}L_{kn}-\delta_{jn}L_{km}-\delta_{km}L_{jn}+\delta_{kn}L_{jm}\bigr)\\
    \commutator{L_{jk}}{\fradkin{mn}}&=\unitimaginary\hbar\bigl(\delta_{jm}\fradkin{kn}+\delta_{jn}\fradkin{km}-\delta_{km}\fradkin{jn}-\delta_{kn}\fradkin{jm}\bigr)\\
    \commutator{\fradkin{jk}}{\fradkin{mn}}&=\frac{\unitimaginary\hbar}{4}\omega^{2}\bigl(\delta_{jm}L_{kn}+\delta_{jn}L_{km}+\delta_{km}L_{jn}+\delta_{kn}L_{jm}\bigr).
\end{split}
\end{equation}
We can also find that
\begin{equation}\label{eqn:orb-frad-ham-rel}
    H^{2}=\smashoperator{\sum_{j,k=1}^{D}}\left(\fradkin{jk}\fradkin{jk}+\frac{\omega^{2}}{4}L_{jk}L_{jk}\right)+\frac{D(D-1)}{4}\hbar^{2}\omega^{2}.
\end{equation}

Introducing,
\begin{equation}
    \definedas{l_{jk}}{\frac{1}{\hbar}L_{jk}},\qquad
    \definedas{\lowfradkin{jk}}{\frac{2}{\hbar\omega}\fradkin{jk}},
\end{equation}
one may easily verify that the Lie algebra $\set{l_{jk},\lowfradkin{mn}}$ under commutator is isomorphic as a Lie algebra to $\complexified*{\unitaryliealgebra*{D}}$ with Hermitian matrix basis,
\begin{equation}
    \rho(l_{jk})_{lm}=-\unitimaginary\delta_{jl}\delta_{km}+\unitimaginary\delta_{jm}\delta_{kl},\qquad
    \rho(\lowfradkin{jk})_{lm}=\delta_{jl}\delta_{km}+\delta_{jm}\delta_{kl}.
\end{equation}
We have therefore identified a hidden $\unitarygroup{D}$ symmetry within the $D$-dimensional isotropic harmonic oscillator problem. Excluding the Hamiltonian from this algebra, we move to a spanning set $\set{l_{jk},b_{mn}}$ for $\complexified*{\specialunitaryliealgebra*{D}}$ defining,
\begin{equation}
    \definedas{b_{jk}}{\lowfradkin{jk}-\frac{1}{D}\delta_{jk}\sum_{m=1}^{D}\lowfradkin{mm}}.
\end{equation}
This yields the commutation relations,
\begin{equation}\label{eqn:su-d-comm-rels}
\begin{split}
    \commutator{l_{jk}}{l_{mn}}&=\unitimaginary\bigl(\delta_{jm}l_{kn}-\delta_{jn}l_{km}-\delta_{km}l_{jn}+\delta_{kn}l_{jm}\bigr)\\
    \commutator{l_{jk}}{b_{mn}}&=\unitimaginary\bigl(\delta_{jm}b_{kn}+\delta_{jn}b_{km}-\delta_{km}b_{jn}-\delta_{kn}b_{jm}\bigr)\\
    \commutator{b_{jk}}{b_{mn}}&=\unitimaginary\bigl(\delta_{jm}l_{kn}+\delta_{jn}l_{km}+\delta_{km}l_{jn}+\delta_{kn}l_{jm}\bigr).
\end{split}
\end{equation}
In this description, \Cref{eqn:orb-frad-ham-rel} becomes,
\begin{equation}
    \frac{D-1}{D}H^{2}=\frac{\hbar^{2}\omega^{2}}{4}\left(\smashoperator[r]{\sum_{j,k=1}^{D}}\bigl(b_{jk}b_{jk}+l_{jk}l_{jk}\bigr)+D(D-1)\right).
\end{equation}
Note: $\set{l_{jk},b_{mn}}$ is a spanning set for $\complexified*{\specialunitaryliealgebra*{D}}$ since
\begin{equation}
    \smashoperator{\sum_{j,k=1}^{D}}\delta_{jk}b_{jk}=0.
\end{equation}
This affords a flexibility in this description, allowing us to render our expressions in a highly symmetric form.

Restricting our discussion to $D=3$, usually one analyses this Lie algebra with respect to the Gell-Mann matrices~\cite{greiner_quantum_1994}. However, the set $\set{l_{jk},b_{mn}}$ renders $\complexified*{\specialunitaryliealgebra*{3}}$ in a particularly symmetric language. The equivalent Casimir elements to those usually defined using a half times the Gell-Mann matrices in terms of our spanning set are
\begin{subequations}
\begin{gather}
    C_{1}=\frac{1}{8}\smashoperator[r]{\sum_{j,k=1}^{3}}\bigl(b_{jk}b_{jk}+l_{jk}l_{jk}\bigr),\\
    C_{2}=-\frac{1}{32}\begin{aligned}[t]
        \bigl(\tensorsym{&b_{11}}{l_{23}}{l_{23}}+\tensorsym{b_{22}}{l_{13}}{l_{13}}+\tensorsym{b_{33}}{l_{12}}{l_{12}}\\
        &-2\,\tensorsym{b_{12}}{l_{13}}{l_{23}}+2\,\tensorsym{b_{13}}{l_{12}}{l_{23}}-2\,\tensorsym{b_{23}}{l_{12}}{l_{13}}\\
        &+\tensorsym{b_{11}}{b_{23}}{b_{23}}+\tensorsym{b_{13}}{b_{13}}{b_{22}}+\tensorsym{b_{12}}{b_{12}}{b_{33}}\\
        &-2\,\tensorsym{b_{12}}{b_{13}}{b_{23}}-\tensorsym{b_{11}}{b_{22}}{b_{33}}\bigr),\label{eqn:cubic-cas}
    \end{aligned}
\end{gather}
\end{subequations}
with
\begin{equation}
    \definedas{\tensorsym{A_{1}}{A_{2}}{A_{3}}}{\smashoperator{\sum_{\sigma\in S_{3}}}A_{\sigma(1)}A_{\sigma(2)}A_{\sigma(3)}}.
\end{equation}
Knowing the form of \Cref{eqn:cubic-cas} in our spanning set, we see that
\begin{equation}
    H^{2}=3\hbar^{2}\omega^{2}\left(C_{1}+\frac{3}{4}\right),\qquad
    H\left(H^{2}-\frac{9\hbar^{2}\omega^{2}}{4}\right)=9\hbar^{3}\omega^{3}C_{2},
\end{equation}
and so,
\begin{subequations}
\begin{gather}
    HC_{1}=3\hbar\omega C_{2}\label{eqn:ham-cas}\\
    C_{2}^{2}=\frac{1}{3}C_{1}^{2}\left(C_{1}+\frac{3}{4}\right)\label{eqn:cubic-quad-rel}
\end{gather}
\end{subequations}

For an energy eigenstate $E$ which is simultaneously an irreducible representation of $\complexified*{\specialunitaryliealgebra*{3}}$ with highest weights $(p,q)$ with $p,q\in\naturals$ (according the convention of Ref.~\cite{pfeifer_lie_2003}), the Casimir elements take on the following values~\cite{pais_dynamical_1966}\footnote{Note: Pais uses a different convention for labelling representations to Ref.~\cite{pfeifer_lie_2003}.}:
\begin{subequations}
\begin{gather}
    C_{1}=\frac{1}{3}(p^{2}+q^{2}+pq+3p+3q),\\
    C_{2}=\frac{1}{18}(p-q)(p+2q+3)(q+2p+3).\label{eqn:cubic-cas-weights}
\end{gather}
\end{subequations}
Note that $C_{1}$ is non-negative. Since $H$ is a real linear combination of the Hermitian $\set{p_{j}^{2},q_{k}^{2}}$ for $j,k\in\set{1,2,3}$, not all of these terms can be simultaneously zero on a state, and so $H$ is positive-definite. Thus, we must have \Cref{eqn:cubic-cas-weights} positive for \Cref{eqn:ham-cas} to hold so $p>q$. Considering \Cref{eqn:cubic-quad-rel} for this representation, we find,
\begin{equation}
    0=-\frac{1}{12}pq(2+p)(2+q)(1+p+q)(3+p+q),
\end{equation}
and so we must have $q=0$ and $p\in\pintegers$. This means that \Cref{eqn:ham-cas} becomes,
\begin{equation}
    \frac{1}{3}p(p+3)E=3\hbar\omega\frac{1}{18}p(p+3)(2p+3),
\end{equation}
and so,
\begin{equation}
    E=(p+\frac{3}{2})\hbar\omega,
\end{equation}
as expected.

\subsection{Discussion}

There are many algebras we could have used to find the spectrum of the harmonic oscillator, with greater complexity affording greater descriptive power. Indeed, Ref.~\cite{dothan_finite-dimensional_1970} uses $\suthreeonec$ to solve this problem, and offers many more examples of the use of spectrum generating algebras, but in our case $\suoneonec$ is the simplest choice. This approach also allows one to relate seemingly quite different physical systems: for example, Ref.~\cite{zeng_most_1994} uses a realisation of $\suoneonec$ for the hydrogen system in $D$-dimensions to relate that system to the $D$-dimensional harmonic oscillator. It is also possible to relate the ladder operator and $\suoneonec$ approaches together, as shown in Ref.~\cite{basu_lorentz_1978}. Our derivation using $\suthreec$ is in complete analogy with the hidden $\sofourc$ derivation we will perform for hydrogen in \Cref{sec:hydrogen}. Each of these methods utilise completely different aspects of the problem to solve it, demonstrating the power and versatility of algebraic approaches.

It is notable that the methods using $\suoneonec$ and $\suthreec$ makes minimal use of the Hilbert space structure of they state space compared to the ladder operator approach. Of course, this reliance has been replaced with a greater dependence on the algebras' representation theories. This is common across the various algebraic methods used in quantum mechanics but remains mysterious next to the clear physical character of the observables we are manipulating. We will see in \Cref{sec:spin-algebras} that this need not always be the case: when considering particle spin in a non-relativistic theory, we can understand the elementary structure of these systems without appealing to Hilbert space or traditional Lie algebra representation theory at all.

\section{Hydrogen}\label{sec:hydrogen}

The theoretical analysis of the hydrogenic atom was a major success of the new quantum mechanics over classical physics. It is an essential problem for any student of physics to solve, and forms the basis for much of spectral theory. Despite this great visibility, solved via differential equations the hydrogenic atom obscures many details of its structure; details that, once understood, render its solution immediate and draw closer parallels with the classical Kepler problem. In this section, we present and then expand on the solution to the hydrogen problem as originally given by Pauli, and indicate how this algebra can be expanded further to begin to account for interactions with hydrogenic systems.

\subsection{Hydrogen Spectrum by Algebras - Pauli}

In Ref.~\cite{pauli_uber_1926}, Pauli gave the first algebraic derivation of the spectrum of hydrogen in three-dimensional Euclidean space using Heisenberg's matrix mechanics. His work was received by Zeitschrift f\"ur Physik five days before Proceedings of the Royal Society of London received Dirac's algebraic solution to the hydrogen problem~\cite{dirac_quantum_1926}, and ten days before Annalen der Physik received \schrodinger's hydrogen spectrum (and wave mechanics) paper~\cite{schrodinger_quantisierung_1926-1}. The algebraic method of both Pauli and Dirac is notable in that we do not have to explicitly find any of the energy eigenstates to derive the spectrum of the Hamiltonian. We will briefly recount Pauli's derivation, as it contains many ideas central to the application of algebras. We will largely follow the presentation given in~\cite{jordan_quantum_1986}\footnote{Not least to avoid the $\mathfrak{Fraktur}$.}, using some notation from~\cite{valent_hydrogen_2003}. Explicit algebraic calculations can be found in~\cite{singer_linearity_2005}, but will not be performed here. Unlike Dirac, whose work we will later briefly recount, Pauli also algebraically calculated the first order perturbation theory for hydrogen in linear electric and magnetic fields; we will not discuss this here, but a more modern discussion can be found in~\cite{valent_hydrogen_2003}. A translation of Pauli's original paper can be found in~\cite{van_der_waerden_sources_1967}

Pauli mirrored an algebraic solution to the classical Kepler problem known since Laplace~\cite{goldstein_prehistory_1975, collas_algebraic_1970}. Unlike Dirac and \schrodinger, it was not necessary for him to write the Hamiltonian in spherical polar coordinates before solving; instead, he related his hydrogenic Hamiltonian\footnote{Strictly, Pauli did not use the reduced mass $\reducedmass$, but it changes nothing to be more precise here.},
\begin{equation}
    H=\frac{1}{2\reducedmass}P^{2}-\potentialcoefficient R^{-1},
\end{equation}
where,
\begin{subequations}
\begin{gather}
    \reducedmass=\frac{1}{m_{\textup{nucl}}}+\frac{1}{m_{e^{-}}}\\
    \potentialcoefficient=\frac{Ze^{2}}{4\pi\varepsilon_{0}}\\
    R^{2}=\threesum{j}Q_{j}^{2},
\end{gather}
\end{subequations}
with two families of conserved operators: the first are the well-known orbital angular momentum operators,
\begin{equation}
    L_{j}=\half\threesum{k,l}\levicivita{j}{k}{l}(Q_{k}P_{l}-P_{k}Q_{l});
\end{equation}
the second are, up to a scalar multiple, quantum mechanical operator versions of the components of the ``Laplace-Runge-Lenz'' vector~\cite{goldstein_classical_2002},
\begin{equation}
    A_{j}=\frac{1}{2\reducedmass\potentialcoefficient}\left(\threesum{k,l}\levicivita{j}{k}{l}(L_{k}P_{l}-P_{k}L_{l})\right)+Q_{j}R^{-1}.
\end{equation}
Indeed, all of these operators are conserved,
\begin{equation}
    [H,A_{j}]=[H,L_{j}]=0,\label{eqn:lrl-orb-ham-comm}
\end{equation}
and can be related to each other by,
\begin{equation}\label{eqn:lrl-orb-ham-relations}
    A^{2}=\frac{2}{\reducedmass\potentialcoefficient^{2}}H\left(\hbar^{2} + L^{2}\right)+1,\qquad
    \threesum{j}L_{j}A_{j}=0.
\end{equation}
The former expression can be viewed as a ``squared eccentricity'' operator: on eigenstates of mutual energy and orbital angular momentum the right-hand side is a constant which is less than one (and non-negative) for bound states ($E<0$), greater than one for unbound states ($E>0$), and equal to one for states with zero energy. The latter expression is the quantum mechanical version of the statement that the classical angular momentum pseudovector and the Laplace-Runge-Lenz vector are orthogonal. Note: both expressions are written entirely in terms of conserved quantities; they capture some structural information intrinsic to the hydrogenic atomic system.

Though each of these families commute with the Hamiltonian, they do not commute with each other,
\begin{subequations}
\begin{gather}
    \commutator{L_{j}}{A_{k}}=\unitimaginary\hbar\threesum{l}\levicivita{j}{k}{l}A_{l}\\
    \commutator{A_{j}}{A_{k}}=-\unitimaginary\hbar\frac{2}{\reducedmass\potentialcoefficient^{2}}H\threesum{l}\levicivita{j}{k}{l}L_{l}.\label{eqn:lrl-commutator}
\end{gather}
\end{subequations}
The presence of the Hamiltonian on the right-hand side of \Cref{eqn:lrl-commutator} means that the Lie algebra under commutator of $\set{L_{j},A_{j}}$ is not closed. In principle, one could consider the Lie algebra of $\set{H^{n}L_{j},H^{n}A_{j}}$ for all $n\in\naturals$ under commutator, but this infinite-dimensional and a priori difficult to work with. Instead, what Pauli and others do~\cite{valent_hydrogen_2003, maclay_dynamical_2020}, is consider $\set{L_{j},A_{j}}$ on a subspace with a given energy $E$, as, by \Cref{eqn:lrl-orb-ham-comm}, their action is closed in this subspace. The principal case of interest is bound states, and so $E<0$. Writing,
\begin{equation}
    K_{j}=\sqrt{\frac{\reducedmass\potentialcoefficient^{2}}{2\abs{E}}}A_{j},
\end{equation}
our now closed six-dimensional Lie algebra is defined by,
\begin{equation}\label{eqn:bound-state-lie-algebra}
    \commutator{L_{j}}{L_{k}}=\unitimaginary\hbar\threesum{l}\levicivita{j}{k}{l}L_{l},\quad
    \commutator{L_{j}}{K_{k}}=\unitimaginary\hbar\threesum{l}\levicivita{j}{k}{l}K_{l},\quad
    \commutator{K_{j}}{K_{k}}=\unitimaginary\hbar\threesum{l}\levicivita{j}{k}{l}L_{l}.
\end{equation}

The route that Pauli took to finding the energy spectrum was to change basis,
\begin{equation}\label{eqn:m-n-basis}
    M_{j}=\frac{1}{2\hbar}(L_{j}+K_{j}),\qquad
    N_{j}=\frac{1}{2\hbar}(L_{j}-K_{j}),
\end{equation}
in which \Cref{eqn:bound-state-lie-algebra} simplifies to,
\begin{equation}
    \commutator{M_{j}}{M_{k}}=\unitimaginary\threesum{l}\levicivita{j}{k}{l}M_{l},\qquad
    \commutator{N_{j}}{N_{k}}=\unitimaginary\threesum{l}\levicivita{j}{k}{l}N_{l},\qquad
    \commutator{M_{j}}{N_{k}}=0,
\end{equation}
and we use \Cref{eqn:lrl-orb-ham-relations} to find,
\begin{equation}
    M^{2}=N^{2},\qquad
    4M^{2}+1 = -\frac{\reducedmass\potentialcoefficient^{2}}{2\hbar^{2}E}.
\end{equation}
At this point, we recognise the $\set{M_{j}}$ Lie subalgebra as $\sutwoc$ and employ knowledge of its representation theory~\cite{fulton_representation_1991}; since $\commutator{H}{M_{j}}=0$, we may restrict our consideration of energy eigenstates to those which are also irreducible representations of $\sutwoc$ generated by the $\set{M_{j}}$, on which,
\begin{equation}
    M^{2}=r(r+1)
\end{equation}
for $r\in\set{0,\half,1,\dots}$. Thus, we find,
\begin{equation}
    E_{n}=-\frac{\reducedmass\potentialcoefficient^{2}}{2\hbar^{2}n^{2}},
\end{equation}
with $n=2r+1\in\pintegers$. We are not quite finished: the $M_{j}$ have no clear physical significance, so should ideally not be used to label our energy eigenstates. Since the $\set{L_{j}}$ form a closed Lie algebra also isomorphic to $\sutwoc$, we attempt to relate the energy eigenstates in terms of $\set{M_{j}}$ with those of $\set{L_{j}}$. Following~\cite{valent_hydrogen_2003} and from \Cref{eqn:m-n-basis} we have,
\begin{equation}\label{eqn:l-in-m-and-n}
    L_{j}=\hbar(M_{j}+N_{j}).
\end{equation}
Considering mutual eigenstates of $\set{H,M_{3},N_{3},L_{3}}$, \Cref{eqn:l-in-m-and-n} implies $-(n-1)\leq m\leq n-1$, where $m\hbar$ is the eigenvalue of $L_{3}$; thus, the representation theory for the $\set{L_{j}}$-generated $\sutwoc$ tells us that $0\leq l\leq n-1$, with $l(l+1)\hbar^{2}$ the eigenvalue of $L^{2}$. We have now recovered the relationship between the principal quantum number $n$ and the maximum orbital angular momentum projection $l$; paired with an algebraic proof that $l\in\naturals$~\cite{ballentine_quantum_2014}, we have completed the derivation for the spectrum of a hydrogenic atom.

\subsection{Hydrogen Spectrum by Algebras - Dirac}

Before expanding our discussion to include representation theory, which will greatly simplify the above argument, it is worth briefly considering Dirac's contribution~\cite{dirac_quantum_1926}. His derivation, while also algebraic, is much more rooted in the theory of canonical variables; in particular, his work utilised quantum mechanical versions of action-angle variables $\set{w_{k},J_{k}}$~\cite{goldstein_classical_2002} to study multiply periodic systems, i.e., those for which their Hamiltonian is dependent only on the generalised momenta $\set{J_{k}}$. He found a quantum version of the classical equation of elliptic orbital motion,
\begin{equation}
    \frac{a}{r}=1+e\cos(\theta-\alpha),
\end{equation}
with $a$ the semi-latus rectum, by writing the Cartesian Kepler Hamiltonian in polar form. He then used his quantum version to find the appropriate action-angle variables for the hydrogen system, and subsequently derived the spectrum.

Dirac's proof is rather technical, and which algebraic approach is more accessible depends heavily on your background. We have given Pauli's derivation more exposure in this paper as it more neatly captures some of the strengths of approaching problems in quantum mechanics through algebras than Dirac's. However, one thing worth noting is that Dirac starts from the hydrogenic Hamiltonian in two-dimensions and reasons that the correct polar form is (in modern notation),
\begin{equation}
    H=\frac{1}{2m}\left(p_{r}^{2}+\frac{L^{2}}{r^{2}}-\frac{\hbar^{2}}{4r^{2}}\right)-\frac{e^{2}}{r},
\end{equation}
which contains an additional contribution to the kinetic energy absent in the na\"ive polar Hamiltonian. This is an artefact of his working in two-dimensions, as the additional term (appropriately scaled) is only present when the dimension of space is not one or three~\cite{adams_algebraic_1994}.

\subsection{The Degeneracy Algebras for Hydrogenic Systems}

In Pauli and Dirac's day, knowledge of Lie groups, Lie algebras, and their representations was limited within the physics community~\cite{maclay_dynamical_2020}. As one might expect, the approach taken by Pauli can be simplified and extended if one knows a little more representation theory, the gift of hindsight leading us more directly to the energy spectrum. In so doing, we will uncover a broader class of solutions to the hydrogen problem than we initially found.

Revisiting \Cref{eqn:bound-state-lie-algebra}, we identify it as the Lie algebra $\sofourc$~\cite{maclay_dynamical_2020}. This Lie algebra is six-dimensional and has two Casimir elements, which in our realisation are,
\begin{equation}
    C_{1}=\frac{1}{\hbar^{2}}(L^{2}+K^{2}),\qquad
    C_{2}=\frac{1}{\hbar^{2}}\threesum{j}L_{j}K_{j},
\end{equation}
i.e., these objects commute with all generators $\set{L_{j},K_{j}}$. The states we are interested in studying transform under unitary irreducible representations of $\sofourc$; these are all finite-dimensional and specified by a pair of numbers $(j_{0},\eta)$ where $2\abs{j_{0}}\in\naturals$ and $\eta = \abs{j_{0}}+n$ for $n\in\pintegers$~\cite{adams_algebraic_1994}. On such an irreducible representation, the Casimir elements become,
\begin{equation}
    C_{1}=j_{0}^{2}+\eta^{2}-1,\qquad
    C_{2}=j_{0}\eta.
\end{equation}
From \Cref{eqn:lrl-orb-ham-relations}, we see directly that for hydrogen $j_{0}=0$ and,
\begin{equation}
    \eta^{2}=n^{2}=-\frac{\reducedmass\potentialcoefficient^{2}}{2\hbar^{2}E},
\end{equation}
and so we directly arrive at the energy spectrum for hydrogen. This derivation is appealing as we did not have to divine some special basis for our Lie algebra to make progress: once we had identified the Lie algebra we were working with, the energy spectrum was immediately determined by its representation theory, up to additional constraints determined by the way the Lie algebra is realised in our system. It also obviates the degeneracy of the energy levels: the irreducible representations of $\sofourc$ have dimension $\eta^{2}-j_{0}^{2}$, which in our case is $n^{2}$, and are further decomposable into integer $j$ representations of $\sothreec$~\cite{adams_algebraic_1994}. Interestingly, this fact is independent of the integer angular momentum constraint imposed by our realisation of $\sothreec$ by $\set{L_{j}}$.

In addition to making the proof of the energy spectrum for bound states more direct, viewing this problem through an algebraic lens also makes clear which classes of solutions are qualitatively different from each other. To see this, recall that the structure of \Cref{eqn:bound-state-lie-algebra} was fixed by restricting our argument to bound states $E<0$. If we instead considered unbound states $E>0$, the Lie algebra of $\set{L_{j},K_{j}}$ would instead be isomorphic to $\sothreeonec$, better known as the complexified Lie algebra of the Lorentz group~\cite{weinberg_quantum_1995}. This Lie algebra also has two Casimir elements,
\begin{equation}
    C_{1}=\frac{1}{\hbar^{2}}(K^{2}-L^{2}),\qquad
    C_{2}=\frac{1}{\hbar^{2}}\threesum{j}L_{j}K_{j},
\end{equation}
and the unitary irreducible representations are again labelled with two numbers $(j'_{0},\eta')$ with,
\begin{equation}
    C_{1}={\eta}^{\prime 2}-j_{0}^{\prime 2}+1,\qquad
    C_{2}=j'_{0}\eta',
\end{equation}
on these representations. However, this time there are no non-trivial finite-dimensional representations of this kind, and we have $2\abs{j'_{0}}\in\naturals$, $\eta\in\reals$, and $[\eta'\neq0]\vee[j_{0}^{\prime}\neq0]$. Again from \Cref{eqn:lrl-orb-ham-relations}, we find,
\begin{equation}
    {\eta}^{\prime 2}-j_{0}^{\prime 2}=\frac{\reducedmass\potentialcoefficient^{2}}{2\hbar^{2}E},\qquad
    j'_{0}\eta'=0,
\end{equation}
on unbound energy eigenstates. Since $E>0$ in this case (and finite), we must have $\eta^{\prime2}>j_{0}^{\prime2}$, so $j'_{0}=0$ and $\eta\neq0$. Therefore, we find,
\begin{equation}
    E(\eta')=\frac{\reducedmass\potentialcoefficient^{2}}{2\hbar^{2}\eta^{\prime2}},
\end{equation}
which is continuous and unbounded from above. Considering the orbital angular momentum structure of these energy eigenstates, we find that $l\in\naturals$ with no upper limit. This derivation recovers key aspects of the theory of Coulomb scattering~\cite{sakuma_self-energy_2012} from a purely algebraic and representation-theoretic viewpoint. In this case, the corresponding analysis using the \schrodinger\space equation is much more complex than for the bound states, and requires the use of confluent hypergeometric functions to describe the wavefunctions of the system~\cite{yafaev_classical_1997}. Furthermore, writing,
\begin{equation}
    p^{2}=\frac{\reducedmass^{2}\potentialcoefficient^{2}}{\hbar^{2}\eta^{\prime2}},
\end{equation}
our approach has identified a characteristic momentum scale for the scattering problem,
\begin{equation}
    p_{0}=\frac{\reducedmass\potentialcoefficient}{\hbar}=\frac{\hbar}{a_{0}},
\end{equation}
where $a_{0}$ is the Bohr radius.

The final case we should consider are the limiting states for ionisation, namely those with $E=0$. In this case, our $\set{K_{j}}$ are ill-defined, so we return to considering $\set{L_{j},A_{j}}$. Now, this Lie algebra is isomorphic to the complexification of the Lie algebra of the Euclidean group~\cite{adams_algebraic_1994} and may be denoted $\sothreezeroonec\cong\complexified*{(\sothree\ltimes\reals^{3})}$~\cite{subag_symmetries_2018}, with $\ltimes$ the Lie algebra semidirect sum~\cite{antoniou_characterization_1991}. Our Casimir elements are now,
\begin{equation}
    C_{1}=A^{2},\qquad
    C_{2}=\frac{1}{\hbar}\threesum{j}L_{j}A_{j},
\end{equation}
and the unitary irreducible representations are again labelled with two numbers $(j''_{0},\eta'')$ with,
\begin{equation}
    C_{1}={\eta}^{\prime\prime 2},\qquad
    C_{2}=j''_{0}\eta'',
\end{equation}
which from \Cref{eqn:lrl-orb-ham-relations} imply,
\begin{equation}
    {\eta}^{\prime\prime 2}=1,\qquad
    j''_{0}\eta''=0,
\end{equation}
and so $j''_{0}=0$ and $\eta''=\pm1$. These two representations are equivalent~\cite{adams_algebraic_1994}, infinite-dimensional, and with all $l\in\naturals$ orbital angular momentum states present. Collectively, we may refer to $\set{\sofourc,\sothreeonec,\sothreezeroonec}$ as the ``degeneracy algebras'' for hydrogenic systems, as their associated Lie groups are referred to as ``degeneracy groups''~\cite{barut_dynamical_1965}.

\subsection{Transitions between Hydrogen Eigenstates}

In the preceding discussion, we restricted ourselves to different classes of energy eigenstates to ensure the Lie algebra of our conserved observables was closed. This necessitated cutting out a fascinating and useful part of the algebraic structure naturally associated with hydrogenic systems. This can be identified with a subalgebra of an infinite-dimensional Kac-Moody loop algebra in all numbers of spatial dimensions (of standard type in even dimensions and of twisted type in odd) with the Hamiltonian as the loop variable~\cite{daboul_dynamical_1994,daboul_hydrogen_1998}. Indeed, this is established using higher dimensional versions of the angular momentum and Laplace-Runge-Lenz vectors, offering a generalisation of the method used by Pauli~\cite{wipf_algebraic_2006}. A full discussion of this structure is beyond the scope of this work, but it is clear that such an algebra contains elements which enable transitions between hydrogen eigenstates.

To approach such algebras, observe that even if we did not restrict ourselves to eigenspaces with a given energy, all the operators we've used thus far cannot map between eigenstates with different energies: they all commute with the Hamiltonian. Therefore, we must expand our algebraic structure to include operators which do not commute with the Hamiltonian --- i.e., are not constant with respect to the evolution of the system. To this end, we may use the $\suoneonec$ Lie algebra introduced in \Cref{eqn:su-one-one-c} and its representation theory, realised in two distinct ways. 

In the first approach, Adams realises this algebra in $D$-dimensions in terms of model specific versions of the usual radial distance $r$ and radial momentum $p_{r}$ operators,
\begin{equation}
    \definedas{\hat{R}_{j}}{\frac{X_{j}}{R}}=\frac{1}{\gamma}r,\qquad
    \definedas{P_{R}}{\half\sum_{j=1}^{D}\bigl(\hat{R}_{j}P_{j}+P_{j}\hat{R}_{j}\bigr)}=\gamma p_{r};
\end{equation}
this realisation is (with $m=\hbar=\frac{e^{2}}{4\pi\varepsilon_{0}}=1$),
\begin{equation}
\begin{split}
    &\sotwoonegen[1]=\half\left(R P_{R}^{2}+\frac{\tau}{R}-R\right),\qquad
    \sotwoonegen[2]=R P_{R},\qquad
    \sotwoonegen[3]=\half\left(R P_{R}^{2}+\frac{\tau}{R}+R\right),\\
    &\qquad\qquad\definedas{\tau}{\frac{1}{4}(D-1)(D-3)+L^{2}},\qquad
    \definedas{\gamma^{2}}{-\frac{1}{2E}},
\end{split}
\end{equation}
with $L^{2}$ as in \Cref{eqn:d-dim-orbital-cas}. This is the same realisation up to a sign as \cite{zeng_most_1994}. 
Using this algebra, and restricting to a simultaneous energy and orbital angular momentum eigenstate satisfying,
\begin{equation}
    L^{2}\ket{E,l}=l(l+D-2)\ket{E,l},
\end{equation}
Adams converts the hydrogenic Hamiltonian to an eigenvalue equation in $\sotwoonegen[3]$,
\begin{equation}
    (\sotwoonegen[3]-\gamma Z)\ket{E,l}=0.
\end{equation}
From the known representation theory of $\suoneonec$ discussed in \Cref{sec:spectrum-generating-algebra}, he immediately solves this eigenvalue equation to find,
\begin{equation}
    E=-\frac{Z^{2}}{2\left(l+\frac{D-1}{2}+\mu\right)^{2}},
\end{equation}
with $\mu\in\naturals$. Identifying the squared term in the denominator with the square of a generalised principal quantum number, we see the same kind of degeneracy present in three dimensions: states of a given orbital angular momentum can have different energies above some minimum, where this minimum is dependent on the orbital angular momentum eigenvalue. It is worth noting that it was our choice of $\gamma^{2}$ with a minus sign which led to the eigenvalue equation being for $\sotwoonegen[3]$, and our description of the bound states; if we had chosen $\gamma^{2}$ with a plus sign, we would have a $\sotwoonegen[1]$ eigenvalue equation, and we would find the continuous spectrum of unbound states. Thus, we see that, though still disconnected, both bound and unbound states of hydrogen can be captured using the same $\suoneonec$ realisation. Adams also gives analyses of the $D$-dimensional Klein-Gordon and Dirac equations in Coulomb potentials using the same $\suoneonec$ machinery, but we will not discuss these here.

Back in three-dimensions, the second approach, following Mart\'inez-Y-Romero~\cite{martinez-y-romero_su1_2005,martinez-y-romero_algebraic_2007} is similar this one insofar as it solves for the energy spectrum given a fixed orbital angular momentum using a realisation of $\suoneonec$. However, it differs markedly in that it does so by coupling together the radial coordinate and \textit{global phase} of the wavefunction. To see how, we define (again with $m=\hbar=\frac{e^{2}}{4\pi\varepsilon_{0}}=1$),
\begin{equation}
    x=\ln(r\sqrt{-2E}),
\end{equation}
under which the radial wavefunction $F(x)$ satisfies,
\begin{equation}
\begin{gathered}
    \left(\odv[2]{}{x}+\frac{2Z}{\sqrt{-2E}}\exp(x)-\exp(2x)-\frac{1}{4}\right)F(x)=l(l+1)F(x)
\end{gathered}
\end{equation}
for a given angular momentum eigenvalue $l$. Introducing a global phase on the transformed radial wavefunction,
\begin{equation}
    \definedas{V(x,\varphi)}{\exp(\unitimaginary\lambda\varphi)F(x)}
\end{equation}
with $\lambda\in\reals$ a constant to be determined, Mart\'inez-Y-Romero realises $\suoneonec$ in terms of ladder operators,
\begin{equation}
    \sotwoonegen[3]=-\unitimaginary\pdv{}{\varphi},\qquad
    \sotwoonegen[\pm]=\unitimaginary\exp(\pm\unitimaginary\varphi)\left(\pdv{}{x}\mp\exp(x)\mp\unitimaginary\pdv{}{\varphi}+\half\right),
\end{equation}
and from this derives the eigenspectrum of hydrogen, with $\lambda$ turning out to be the principal quantum number. While this approach may seem strange at first, the global phase being typically neglected in quantum mechanics, the authors believe this approach worthy of further consideration.

\subsection{Dynamical Algebras}

In our analysis so far, the degeneracy algebras associated with the hydrogen spectrum have remained separate, both from each other and from the transitional algebraic structure discussed in the previous section; this is a somewhat unnatural state of affairs and motivates the study of algebraic structures which unify all of these into a single framework. This unification occurs within what is known as a ``dynamical group''. A dynamical group of a quantum system is one for which all the states of that system are contained within a single irreducible representation of that group~\cite{barut_calculation_1967}. We term a ``dynamical algebra'' as the Lie algebra of a dynamical group. Intuitively, the operators of a dynamical algebra can be generators of symmetries within a class of states (i.e., degeneracy algebra elements), or can generate transitions between these classes.

The loop Lie algebra of Ref.~\cite{daboul_hydrogen_1998} is an infinite-dimensional example. An important finite-dimensional example of a dynamical algebra is the Lie algebra $\sofourtwoc$ which includes each of the degeneracy algebras $\set{\sofourc,\sothreeonec,\sothreezeroonec}$ and the algebra $\sotwoone\cong\suoneone$ as subalgebras. Notably, both the electromagnetic dipole operator and the \schrodinger\space equation for the system can be expressed using the elements of $\sofourtwoc$~\cite{barut_transition_1967, maclay_dynamical_2020,kleinert_group_1968,adams_algebraic_1994}. Clearly, dynamical algebras offer a compelling encapsulation of important features of the physical system, but are a topic far too expansive for us to consider further here.

\section{Optics}\label{sec:optics}

What separates physics and mathematics is the need to describe a physical system in principle realisable in the real world. The discussion on algebraic methods up to now has aided in this aim, but the significance of the species of algebra we worked with on the physics of the system has been somewhat obscure. In this section, we will see how one may directly model components of optical systems, common to labs the world over, by algebraic means. In so doing, we will see the phenomenological consequences of the algebras we employ in a tangible form.

\subsection{The Jordan-Schwinger Map}\label{sec:jordan-wigner-map}

At the core of the success of quantum mechanics and quantum field theory is the ability to leverage the symmetries present in a system to understand its structure and evolution through unitary representations. An invaluable tool for working with unitary representations of a given Lie algebra, especially in quantum optics, is the ``Jordan-Schwinger map'', named for the method's originator and populariser respectively \cite{jordan_zusammenhang_1935,schwinger_angular_1952}. The method utilises a set of $D$ ladder operators $\set{a_{j},a^{\dagger}_{j}}$, either bosonic,
\begin{equation}\label{eqn:bosonic-ladder-operators}
    \commutator{a_{j}}{a_{k}}=\commutator{a^{\dagger}_{j}}{a^{\dagger}_{k}}=0,\qquad
    \commutator{a_{j}}{a^{\dagger}_{k}}=\delta_{jk},
\end{equation}
or fermionic,
\begin{equation}
    \anticommutator{a_{j}}{a_{k}}=\anticommutator{a^{\dagger}_{j}}{a^{\dagger}_{k}}=0,\qquad
    \anticommutator{a_{j}}{a^{\dagger}_{k}}=\delta_{jk},
\end{equation}
which act on the natural Fock space of mutual number operator eigenstates,
\begin{equation}
    N_{j}\ket{n_{1}\dots n_{j}\dots n_{D}}=a^{\dagger}_{j}a_{j}\ket{n_{1}\dots n_{j}\dots n_{D}}=n_{j}\ket{n_{1}\dots n_{j}\dots n_{D}}.
\end{equation}
Using these operators, we may map any matrix Lie algebra $\set{M_{jk}}$ with commutator as the Lie product to an operator algebra on this Fock space in many ways. For example,
\begin{equation}\label{eqn:standard-jordan-schwinger}
    \mapdefinition{M_{jk}}{\sum_{j,k}M_{jk}a^{\dagger}_{j}a_{k}},
\end{equation}
may easily be proven to be a Lie algebra isomorphism which preserves the Hermiticity of the matrices. Mapping matrices to Fock space operators allows us to use combinations of the $\set{a^{\dagger}_{j}}$ acting on the state $\definedas{\ket{0}}{\ket{0\dots0}}$ characterised by,
\begin{equation}
    a_{j}\!\ket{0}=0
\end{equation}
to directly realise (possibly infinite-dimensional) representations for this Lie algebra. Whether these representations are unitary or not depends on the particular realisation we choose; we will see a non-trivial example of this in our discussion of non-linear optical components. Of course, the Jordan-Schwinger map is more widely applicable than what we will discuss --- see~\cite{lipkin_lie_2003} for many examples in early particle physics, including $\suthreec$ flavour symmetry.

\subsection{Two mode \texorpdfstring{$\sutwo$}{su(2)} Optical Components}

Optical interference experiments, both classical and quantum, are ubiquitous in physics and underpin some of its most important discoveries. Essential to the toolkit for performing such experiments are beam splitters and devices which can alter the phase difference between beams. Suitably interpreted, both beam splitters and phase shifters can be modelled as linear maps which take in a pair of beams as input and output a new pair of beams. We will focus our analysis on the algebraic description of these components, as thorough treatments on optical interference experiments using these components are widely available.

In the case of beams of photons, we may readily model them in the language of Fock space with bosonic ladder operators of the kind defined in \Cref{eqn:bosonic-ladder-operators}. To each pair of beams, we may associate a realisation of $\utwoc$ using the standard Jordan-Wigner map,
\begin{equation}
    I=a^{\dagger}_{1}a_{1}+a^{\dagger}_{2}a_{2},\quad
    J_{x}=\half\left(a^{\dagger}_{1}a_{2}+a^{\dagger}_{2}a_{1}\right),\quad
    J_{y}=-\half[\unitimaginary]\left(a^{\dagger}_{1}a_{2}-a^{\dagger}_{2}a_{1}\right),\quad
    J_{z}=\half\left(a^{\dagger}_{1}a_{1}-a^{\dagger}_{2}a_{2}\right),
\end{equation}
with $\set{J_{x},J_{y},J_{z}}$ the usual basis for $\sutwoc$. In this setting, the annihilation and creation operators remove/add a photon from the quantum state respectively. The action of a linear optical component is given by a 2-by-2 unitary matrix $U$ such that,
\begin{equation}\label{eqn:su2-component}
    \begin{+pmatrix}
        \aout[1]\\
        \aout[2]
    \end{+pmatrix}
    = U\begin{+pmatrix}
        \ain[1]\\
        \ain[2]
    \end{+pmatrix},
\end{equation}
with $\ain[j]$ the annihilation operator for the incoming beam $j$ etc. For an idealised beam splitter, the relevant unitary is~\cite{leonhardt_essential_2010},
\begin{equation}\label{eqn:ideal-beam-splitter}
    U_{BS}(\rho)=\begin{+pmatrix}
        \cos\left(\half[\rho]\right) & -\sin\left(\half[\rho]\right) \\
        \sin\left(\half[\rho]\right) & \cos\left(\half[\rho]\right)
    \end{+pmatrix},
\end{equation}
with $\rho\in\reals$ related to the reflectivity of the beam splitter. We may arbitrarily change the phase relationship between the two beams using,
\begin{equation}\label{eqn:phase-change-unitary}
    U_{PC}(\alpha,\beta)=\begin{+pmatrix}
        \exp(\unitimaginary\alpha) & 0 \\
        0 & \exp(\unitimaginary\beta),
    \end{+pmatrix}
\end{equation}
which can be understood physically as changing the optical path lengths for each beam independently. Utilising these components, we may build any $\sutwoc$ interferometer.

\subsection{The Hong-Ou-Mandel Effect}

The ladder operator formalism is particularly useful as it naturally encodes the exchange statistics of photons. This becomes significant when two photons are identical in their properties and coincident on a beam splitter through the ``Hong-Ou-Mandel effect'' \cite{bouchard_two-photon_2020,Hance_2025}, where the photons ``bunch'' in the output beams. To assess this situation, first consider that the input state to the beam splitter in this situation is,
\begin{equation}
    \ket{\psi}=\adaggerin[1]\adaggerin[2]\ket{0}.
\end{equation}
Hermitian conjugating \Cref{eqn:su2-component} and recalling that \Cref{eqn:ideal-beam-splitter} is unitary we find,
\begin{equation}
    \adaggerin[1]=\cos\left(\half[\rho]\right)\adaggerout[1]+\sin\left(\half[\rho]\right)\adaggerout[2],\quad
    \adaggerin[2]=-\sin\left(\half[\rho]\right)\adaggerout[1]+\cos\left(\half[\rho]\right)\adaggerout[2].
\end{equation}
Thus, on substituting,
\begin{equation}
    \ket{\psi}=\left(\cos(\rho)\adaggerout[1]\adaggerout[2]-\half\sin(\rho)\left(\left(\adaggerout[1]\right)^{2}-\left(\adaggerout[2]\right)^{2}\right)\right)\ket{0}.
\end{equation}
For a 50:50 beam splitter $\rho=\half[\pi]$ and we find,
\begin{equation}
    \ket{\psi}=-\half\left(\left(\adaggerout[1]\right)^{2}-\left(\adaggerout[2]\right)^{2}\right)\ket{0},
\end{equation}
i.e., the probability that the two photons emerge from the beam splitter in different beams is zero.

\subsection{Two mode \texorpdfstring{$\suoneone$}{su(1,1)} Optical Components}

The optical components so far discussed have been ``linear'' devices; classically this means that the polarisation density of the dielectric medium which comprises them is linear in the electric field. To go beyond the limitations of these components, it has become increasingly important to utilise materials for which the polarisation density includes non-negligible terms which are non-linear in the electric field. An important class of processes arising from one of these non-linearities are ``three-wave mixing'' phenomena and in particular ``spontaneous parametric down-conversion''~\cite{couteau_spontaneous_2018} and the closely related ``parametric amplification''. This is a process by which ``pump'' photons input to the material cause the production of pairs of lower frequency photons.

The full theoretical treatment of non-linear optical effects in both classical and quantum physics is beyond the scope of this work, but the relevant details for parametric down-conversion can be found in~\cite{louisell_quantum_1961,mollow_quantum_1967,mollow_quantum_1967-1}. Classically, the non-linear response of the medium causes the amplitudes and phases of the input electromagnetic waves to become coupled. In particular, the evolution of the complex amplitude of one wave is dependent on the conjugate amplitude of another~\cite{kumar_squeezed-state_1984,new_introduction_2011}. Quantum mechanically, this results in superposition between the creation and annihilation operators of the two photon modes, which we may generally parametrise as~\cite{leonhardt_essential_2010},
\begin{equation}
    \begin{+pmatrix}
        \aout[1] \\
        \adaggerout[2]
    \end{+pmatrix}
    = V
    \begin{+pmatrix}
       \ain[1] \\
       \adaggerin[2] 
    \end{+pmatrix}
\end{equation}
with $V$ a 2-by-2 matrix. If we now impose the commutation relations for ladder operators to the output operators, we find that $V\in\uoneonegroup$, i.e., $V$ is an invertible operator preserving the indefinite sesquilinear form defined on $\complexnumbers^{2}$ with signature $(+,-)$. $\uoneonegroup$ is a Lie group whose Lie algebra $\uoneone$ is real and has four generators,
\begin{equation}    
    \sotwoonegen[0]'=\begin{+pmatrix}
        \unitimaginary & 0 \\
        0 & \unitimaginary
    \end{+pmatrix},\quad
    \sotwoonegen[1]'=\half\begin{+pmatrix}
        0 & 1 \\
        1 & 0
    \end{+pmatrix},\quad
    \sotwoonegen[2]'=\half\begin{+pmatrix}
        0 & \unitimaginary \\
        -\unitimaginary & 0
    \end{+pmatrix},\quad
    \sotwoonegen[3]'=\half\begin{+pmatrix}
        \unitimaginary & 0 \\
        0 & -\unitimaginary
    \end{+pmatrix}.
\end{equation}
We identify the Lie subalgebra $\set{\sotwoonegen[1]',\sotwoonegen[2]',\sotwoonegen[3]'}$ under commutator as a representation of the real Lie algebra $\suoneone$.

If we were to use \Cref{eqn:standard-jordan-schwinger} to realise $\uoneone$ on the Fock space, we immediately encounter a problem. Since the matrix generators $\set{\sotwoonegen[j]'}$ are not all anti-Hermitian with respect to the usual Hermitian conjugation (conjugate transpose), their exponentials will not be unitary transformations, thus the representation not unitary. To overcome this obstacle, an alternative realisation must be used in which the generators have the same Hermiticity. Details of this method can be found in~\cite{gromov_jordan-schwinger_1990,gromov_jordan-schwinger_1990-1}. For consistency with our discussion of $\sutwoc$, we give a Hermitian basis for the complexification realised on two modes~\cite{holman_complex_1966},
\begin{equation}
\begin{split}
    \sotwoonegen[0]&=a^{\dagger}_{1}a_{1}-a_{2}a^{\dagger}_{2},\qquad
    \sotwoonegen[1]=\half\left(a^{\dagger}_{1}a^{\dagger}_{2}+a_{1}a_{2}\right),\\
    \sotwoonegen[2]&=-\half[\unitimaginary]\left(a^{\dagger}_{1}a^{\dagger}_{2}-a_{1}a_{2}\right),\qquad
    \sotwoonegen[3]=\half\left(a^{\dagger}_{1}a_{1}+a_{2}a^{\dagger}_{2}\right).
\end{split}
\end{equation}
Realisations for other numbers of modes are possible and used~\cite{hong-yi_new_1992,yurke_su2_1986}. The relevant transformation for an idealised non-linear crystal within an $\suoneonec$ interferometer is,
\begin{equation}
V_{NLC}(\beta)=\begin{+pmatrix}
    \cosh\left(\half[\beta]\right) & -\unitimaginary\sinh\left(\half[\beta]\right)\\
    \unitimaginary\sinh\left(\half[\beta]\right) & \cosh\left(\half[\beta]\right)
\end{+pmatrix},
\end{equation}
with $\beta\in\reals$. Together with arbitrary phase changes in each arm i.e., \Cref{eqn:phase-change-unitary}, we may realise any $\suoneonec$ interferometer.

\subsection{Imaging with Undetected Photons}

The advent of $\suoneone$ interferometry~\cite{yurke_su2_1986} brought with it a method to directly leverage the quantum properties of light to achieve things impossible in classical physics.

\begin{figure}[H]
    \centering
    \includegraphics[width=\linewidth]{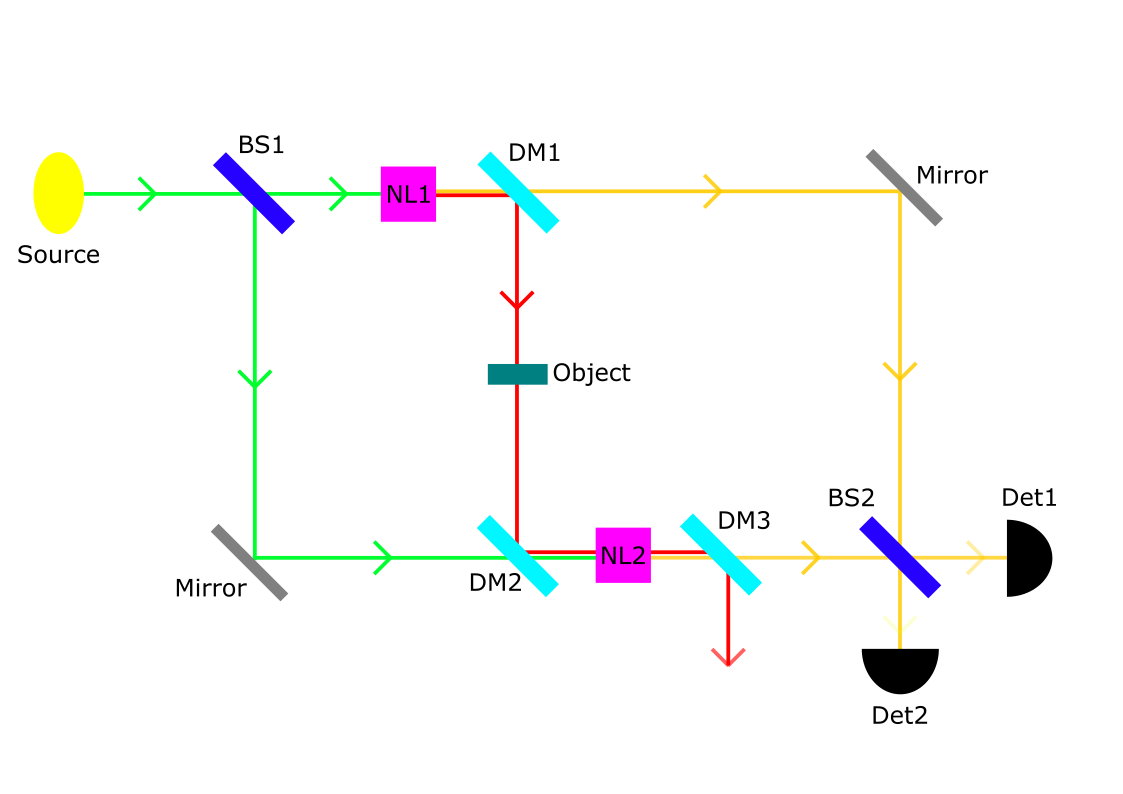}
    \caption{Lemos \textit{et al.}'s protocol for imaging with undetected photons, a form of SU(1,1) interferometry~\cite{lemos_quantum_2014}. In reality, the pump beams (green) are not depleted by the non-linear crystals (NL), but are omitted from the rest of the diagram for simplicity. See text for description.}
    \label{fig:ImagingwUndetectedPhotons}
\end{figure}

A particularly striking example of this is Lemos et al.'s protocol for ``Imaging with Undetected Photons''~\cite{lemos_quantum_2014} which uses quantum entanglement to transfer phase information from a photon which interacts with an Object to another which is then interfered. This setup is shown in \Cref{fig:ImagingwUndetectedPhotons}. Here, a pump beam (green) is emitted by the source, and split into two by a 50:50 beam-splitter (BS1). The transmitted output of BS1 pumps a nonlinear crystal (NL1), potentially leading to the pair creation of signal (yellow) and idler (photons) through spontaneous parametric down-conversion (SPDC). (The pump beam on this path is then filtered away, by e.g., a low-pass filter.) The dichroic mirror DM1 then transmits the signal beam (which continues on to 50:50 beam-splitter BS2) while reflecting the idler beam down towards the Object under investigation. If the idler photon is not absorbed by the Object, it continues down to be reflected by dichroic mirror DM2 into joining the reflected part of the pump beam in entering nonlinear crystal NL2.

Due to the equal probabilities of down-conversion in NL1 and NL2, and the vanishingly small probability of simultaneous down-conversion in both, assuming the Object does not block the idler beam's path, the idler beam exiting NL2 and being separated off by dichroic mirror DM3 carries no information as to whether it was generated at NL1 or NL2; the ``which-down-conversion'' information is erased. However, if the Object would have blocked the idler photon's path, any idler photons exiting NL2 must have been emitted at NL2. Even without then detecting the idler beam separated out by DM3, this presence or absence of ``which-down-conversion'' information affects the interference between the two signal beams (one from NL1, one from NL2) at the beam-splitter BS2. Due to how the paths are balanced, if there is no Object blocking the idler path, then the signal photon will always arrive in detector D2. This means, if we see a detection in detector D1 (which we can herald with the idler photon from DM3 to ensure it isn't a dark count), then the Object must have been present. This is analogous to how which-path information erasure/visibility allows interaction-free measurement in $\sutwo$ interferometry in the Elitzur-Vaidman Bomb Tester~\cite{elitzur_quantum_1993}. Additional information about both $\sutwo$ and $\suoneone$ interferometry is available in~\cite{ou_quantum_2020}.

\subsection{Discussion}

This section has aimed to closely link the quantum nature of light with the algebra of bosonic creation and annihilation operators, and show how common optical components may be readily modelled using this language. This has been done at a high level to avoid the inevitable tangling in the weeds that occurs when discussing such a rich and diverse area. However, it should be stated that one can approach these topics with a much higher degree of algebraic reasoning, arriving at the same conclusions but much more directly and generally. Such an approach will be expounded in future work by the authors.

\section{Spin of Non-Relativistic Systems}\label{sec:spin-algebras}

Since its discovery in the 1920s, the phenomenon of particle spin has become ubiquitous in physics. As such, there are myriad ways to describe it mathematically. The standard approach through representation theory is well-known: we may model a non-relativistic particle with a given spin $s\in\set{0,\half,1,\half[3]\dots}$ using an irreducible representation of the Lie group $\specialunitarygroup{2}$. We will not recount the details here, but a treatment of the general method can be found in many standard texts~\cite{zare_angular_1988,binney_physics_2013,hall_lie_2015,humphreys_introduction_1972,varadarajan_lie_1984}. Complementary to the standard treatment are many others which capture the same mathematical ideas, which we will explore in this section. Quite compellingly, we will see that as the approaches become more algebraic, more of the physical structure of these systems is revealed.

\subsection{Ladder Operators}

The physicists favourite workhorse makes a return in the story of spin, as the spin degrees of freedom of all non-relativistic systems may be accessed using ladder operators. This construction is due to Schwinger~\cite{schwinger_angular_1952, sakurai_modern_2020}, and uses the Jordan-Schwinger map previously discussed in \Cref{sec:jordan-wigner-map}. Taking two commuting species of bosonic creation and annihilation operators $\set{a_{\pm},a^{\dagger}_{\pm}}$ we may realise $\sutwoc$ using the usual ladder operator basis,
\begin{equation}
\begin{gathered}
    \sothreegen[\pm]=\hbar a^{\dagger}_{\pm}a_{\mp}\\
    \sothreegen[z]=\half[\hbar]\left(a^{\dagger}_{+}a_{+}-a^{\dagger}_{-}a_{-}\right).
\end{gathered}
\end{equation}
From this, it is easy to verify that an arbitrary spin state can be realised in the shared bosonic Fock space as,
\begin{equation}
    \ket{s,m}=\frac{(a^{\dagger}_{+})^{s+m}(a^{\dagger}_{-})^{s-m}}{\sqrt{(s+m)!(s-m)!}}\ket{0},
\end{equation}
with,
\begin{subequations}
\begin{gather}
    \sothreecasimirelement\ket{s,m}=s(s+1)\ket{s,m}\\
    \sothreegen[z]\ket{s,m}=m\ket{s,m}
\end{gather}
\end{subequations}
and $a_{\pm}\ket{0}=0$ as usual. If we restrict our access to the state space to a particular value of $s$, the resulting vector space is naturally an irreducible representation of $\sutwoc$ with spin $s$. Extensions of this method to a larger number of oscillator modes, realising the generators of $\sutwoc$ in terms of a higher spin representation, are also used~\cite{dubus_bosons_2024}.

This model for states has great utility, and can be generalised to represent irreducible representations of other Lie algebras~\cite{iachello_lie_2015}. Despite its simplicity, it already demonstrates that most of the important structure contained in the state space for the spin system is ultimately derived from the algebraic structure underlying the theory. It is spiritually related to Dirac's ``standard ket''~\cite{dirac_principles_1981}, a state from which one may use a subset of operators to reach any other state.

In physical terms, the realisation of arbitrary spin states in this way can be interpreted as particular configurations of spin-$\half$ particles; this reflects the fact that the spin-$\half$ representation is a ``fundamental'' representation. While mathematically consistent, applying this method to hypothetical elementary particles with arbitrary spin is physically unsatisfying. What physical status should the underlying oscillators be given in this case? This observation motivates a different approach which does not rely on unphysical substructure.

\subsection{Spherical Tensor Operators}

A strongly algebraic model which avoids questions of unphysicality is Racah's ``spherical tensor'' formalism~\cite{thompson_angular_2008,biedenharn_angular_1984}. This is an approach to handling transformations of systems with large angular momenta systematically, especially composite systems. For our purposes, they model the algebra of operators for a spin-$s$ system, where each family of ``spherical tensor operators'' are themselves irreducible representations of $\sutwoc$. More precisely, we define a set of operators $\set{T(k,q)\mid\forall k\in\naturals,q\in\integers,\abs{q}\leq k}$, which transform irreducibly under the adjoint action of $\sutwoc$,
\begin{subequations}
\begin{gather}
    \commutator{S_{z}}{T(k,q)}=\hbar q\,T(k,q)\\
    \commutator{S_{\pm}}{T(k,q)} = \hbar\sqrt{k(k+1)-q(q\pm1)}\, T(k,q\pm1).\label{eqn:ladder-operator-spherical-tensor}
\end{gather}
\end{subequations}
This may be done constructively defining,
\begin{subequations}
\begin{gather}
    T(0,0)=I_{2s+1}\\
    T(q, \pm q)=\left(\mp\frac{1}{\sqrt{2}}S_{\pm}\right)^{q}
\end{gather}
\end{subequations}
with $I_{2s+1}$ the identity operator, and then using \Cref{eqn:ladder-operator-spherical-tensor} to define the $T(q,k)$ for $\abs{k}<q$. The applications of this formalism are widespread, but we will focus instead on an easily overlooked aspect of spherical tensor operators: their physicality.

It is clear that only $T(q,0)$ are Hermitian, but $T(q,\pm\abs{k})$ may be linearly combined to produce Hermitian operators. The total dimension of independent Hermitian operators entailed by a system of spherical tensor operators is $(2s+1)^{2}$, and thus span the Hermitian operators for the spin degrees of freedom of the spin-$s$ system. This basis for Hermitian operators has a natural and hierarchical relationship with $\specialunitarygroup{2}$, and thus $\specialorthogonalgroup{3}$ rotational symmetry. Each family of spherical tensor operators transforms with respect to a different irreducible representation of $\specialunitarygroup{2}$, specifically as a particle with angular momentum $q\hbar$. As such, each family is both mathematically and \textit{physically} distinguishable. That the totality of these families of operators spans the Hermitian operators, we are empowered to describe operators in a language which naturally relates to physically meaningful concepts, a highly desirable property when constructing a physical model.

Conceptually, this is an improvement over the ladder operator approach as it avoids much of the latter's unphysical features. However, the spherical tensor formalism is not perfect in this regard. The first thing to note is their definition requires access to the standard spin ladder operators $\sothreegen[\pm]$. While this is not usually controversial in physics, it is important to note that these operators exist only within the \textit{complexification} $\sutwoc$ of the real Lie algebra $\sutwo$, and that exponentiations of $\sutwoc$ elements are generically \textit{not} elements of $\specialunitarygroup{2}$. As such, we have moved away from discussing the system in terms of the symmetry we started with. Moreover, the complex numbers we have introduced lack a clear physical interpretation and origin. This is often justified physically by saying that quantum mechanics requires complex numbers, and so this is harmless. However, the origin of complex numbers in quantum mechanics is itself poorly understood in physical terms, ad theories such as Koopman-von Neumann classical physics demonstrates that the role of complex numbers in interference phenomena is typically overstated in the literature~\cite{bondar_operational_2012}. Another limitation of this formalism is the requirement to define a quantisation axis with respect to which we define the spherical tensor operators, essential here as each operator is an eigenstate of the adjoint action of $\sothreegen[z]$. This is a common requirement with the usual higher-spin matrix formalism~\cite{binney_physics_2013} (which also requires a choice of eigenstate basis); given the description of this phenomenon began with a \textit{symmetry}, the need to explicitly \textit{break} this symmetry in order to describe its representations feels somewhat unnatural.

\subsection{Real Algebraic Approach}

To overcome the remaining conceptual difficulties of the spherical tensor formalism, we must adopt a much more mathematically frugal approach. Is it possible to access the underlying algebraic structure of the spin of non-relativistic systems without requiring a quantisation axis or complexification? We will overview an approach to this question detailed (more carefully than we will be here) inRef.~\cite{bradshaw_algebraic_2024}, as the methods required to do so are wholly (real) algebraic in nature and do not rely on the existence of eigenstates for the generators of $\sutwo$.

Let us begin with the real Lie algebra $\sutwo$ which we may define using three generators $\set{\sothreegen[j]}$ and Lie product,
\begin{equation}
    \sothreelieproductdefinition{a}{b}{c}.
\end{equation}
The most general real associative algebra with an identity element in which $\sutwo$ is a subspace is the tensor algebra $\tsutwo$. We may embed the structure of $\sutwo$'s Lie product into the commutator of $\tsutwo$ by quotienting it by an appropriate two-sided ideal generated by all tensors of the form $\tensor{x;y}-\tensor{y;x}-\sothreelieproduct{x}{y}$ with $x,y\in\sutwo$,
\begin{equation}
    \usutwo\cong\frac{\tsutwo}{\ideal{\tensor{x;y}-\tensor{y;x}-\sothreelieproduct{x}{y}}}.
\end{equation}
Doing so forms the universal enveloping algebra $\usutwo$.

As $\usutwo$ is infinite-dimensional, our goal is to quotient it further to form finite-dimensional algebras which preserve the Lie algebra structure of $\sutwo$ found within. To do this, we study the relationship between the adjoint action of the Casimir element,
\begin{subequations}
\begin{gather}
    \definedas{\sothreecasimirelement}{\sothreesum{a}\sothreegen[a]\sothreegen[a]}\\
    \definedas{\sothreead[\sothreecasimirelement]}{\mapdefinition{A}{\sothreesum{a}\bigl[\sothreegen[a],[\sothreegen[a],A]\bigr]}},
\end{gather}
\end{subequations}
with $\commutator{\cdot}{\cdot}$ the usual commutator, and the action of left multiplication,
\begin{equation}
    \definedas{\sothreeleft[A]}{\mapdefinition{B}{AB}},
\end{equation}
where the product of $\usutwo$ has been notated by symbol juxtaposition. From this, one finds $\forall x\in\sutwo$,
\begin{equation}\label{eqn:usutwo-cas-left-identity}
    \bigl[\sothreead[\sothreecasimirelement],\bigl[\sothreead[\sothreecasimirelement],\bigl[\sothreead[\sothreecasimirelement],\sothreeleft[x]\bigr]\bigr]\bigr]+2\bigl[\composition{\sothreead[\sothreecasimirelement]}{\sothreead[\sothreecasimirelement]},\sothreeleft[x]\bigr]=0,
\end{equation}
with $\,\composition{}{}\,$ function composition and $0$ the zero map. Though \Cref{eqn:usutwo-cas-left-identity} looks complicated at first glance, it is actually remarkably simple considering it holds true when acting on \textit{any} element of $\usutwo$, for any given element of $\sutwo$.

Using \Cref{eqn:usutwo-cas-left-identity} we may decompose subspaces of $\usutwo$ into subspaces closed with respect to commutators with elements of $\sutwo$; this decomposition in some sense respects the Lie algebra structure encoded within the commutator of the algebra. To do this, we denote by $\Pi_{k}$ a projection operator into an eigenspace of $\casaction$ with eigenvalue $-k(k+1)$ for $k\in\naturals$, i.e.,
\begin{equation}
    \composition{\casaction}{\Pi_{k}}=-k(k+1)\Pi_{k};
\end{equation}
applying $\Pi_{k}$ to both sides of \Cref{eqn:usutwo-cas-left-identity} reveals,
\begin{equation}
    \composition{\Bigl(\casaction+k(k-1)\id\Bigr)}{\Bigl(\casaction+k(k+1)\id\Bigr)}{\Bigl(\casaction+(k+1)(k+2)\id\Bigr)}{\sothreeleft[x]}{\Pi_{k}}=0.
\end{equation}
This means that on $\composition{\sothreeleft[x]}{\Pi_{k}}$ the operator $\casaction$ has an annihilating polynomial. It turns out that for $k\in\pintegers$, this annihilating polynomial is minimal for all possible $\Pi_{k\in\pintegers}$, and so $\composition{\sothreeleft[x]}{\Pi_{k}}$ decomposes into three trivially-intersecting subspaces,
\begin{equation}
    \composition{\sothreeleft[x]}\Pi_{k}=\Pi'_{k-1}+\Pi'_{k}+\Pi'_{k+1},
\end{equation}
with,
\begin{equation}\label{eqn:k-pinteger-decomp}
    \Pi'_{j}=\begin{dcases}
        \displaystyle\composition{\frac{\composition{\Bigl(\casaction+k(k+1)\id\Bigr)}{\Bigl(\casaction+(k+1)(k+2)\id\Bigr)}}{4k(2k+1)}}{\sothreeleft[x]}{\Pi_{k}} & j=k-1 \\[0.25\baselineskip]
        \displaystyle\composition{\frac{\composition{\Bigl(\casaction+k(k-1)\id\Bigr)}{\Bigl(\casaction+(k+1)(k+2)\id\Bigr)}}{-4k(k+1)}}{\sothreeleft[x]}{\Pi_{k}} & j=k \\[0.25\baselineskip]
        \displaystyle\composition{\frac{\composition{\Bigl(\casaction+k(k-1)\id\Bigr)}{\Bigl(\casaction+k(k+1)\id\Bigr)}}{4(k+1)(2k+1)}}{\sothreeleft[x]}{\Pi_{k}} & j=k+1
    \end{dcases}
\end{equation}
For $k=0$ this annihilating polynomial is not minimal, and in fact for all possible $\Pi_{0}$,
\begin{equation}\label{eqn:k-zero-decomp}
    \composition{\casaction}{\sothreeleft[x]}{\Pi_{0}}=-2\composition{\sothreeleft[x]}{\Pi_{k}}.
\end{equation}

Not only does $\casaction$ have an eigenspace with eigenvalue $-k(k+1)$ for some $k\in\integers$, it has only such eigenspaces. To see this, first note that $\reals$ is an eigenspace of $\casaction$ with eigenvalue $0=-0(0+1)$. Thus, our decomposition begins with an eigenspace of the required form and so by \Cref{eqn:k-pinteger-decomp} and \Cref{eqn:k-zero-decomp} alternating left multiplication by elements of $\sutwo$ with decomposition by $\casaction$, we may resolve an arbitrary element of $\usutwo$ into terms which all inhabit eigenspaces of the required form. Of particular importance are the elements which first attain the eigenvalue $-k(k+1)$ for some particular $k\in\naturals$ during this process. These are called the ``multipoles'', and may be defined recursively as the images of,
\begin{equation}
    \begin{aligned}\label{eqn:recursivestepmultipole}
        \multipole{0}(\alpha)&=\alpha\\
        \multipole{k+1}(xB)&=\composition{\frac{\composition{\Bigl(\casaction+k(k-1)\id\Bigr)}{\Bigl(\casaction+k(k+1)\id\Bigr)}}{4(k+1)(2k+1)}}{\sothreeleft[x]}{\multipole{k}(B)},
    \end{aligned}
\end{equation}
for all $\alpha\in\reals$, $x\in\sutwo$, and $B\in\tensorpower{\sutwo}{k}\subset\usutwo$. Denoting,
\begin{subequations}
\begin{align}
        \definedas{\multipoletensor{a_{1}a_{2}\dots a_{k}}}{&\multipole{k}\bigl(\sothreegen[a_{1}]\sothreegen[a_{2}]\dots\sothreegen[a_{k}]\bigr)},\quad k\in\pintegers\\
        \definedas{\multipoletensor{}}{&\multipole{0}(1)},
\end{align}
\end{subequations}
we give the form of the first few multipoles in \Cref{tab:multipole-images}.

\begin{table}[!t]
\begin{center}
    \makebox[0cm]{
        
\begin{math}
\begin{tblr}{colspec={c | c},rowsep=0.5\baselineskip,hline{2-Y}}
    \textup{Multipole} & \textup{Explicit Form}\\
    \multipoletensor{} & \sothreemonopoledefinition \\
    \multipoletensor{a} & \sothreedipoledefinition{a} \\
    \multipoletensor{ab} & \displaystyle\half\bigl(\generator[a]\generator[b]+\generator[b]\generator[a]\bigr)-\frac{1}{3}\delta_{ab}\sothreecasimirelement \\
    \multipoletensor{abc} & \displaystyle\begin{aligned}
        \frac{1}{6}\bigl(\generator[a]\generator[b]\generator[c]+&\generator[a]\generator[c]\generator[b]+\generator[b]\generator[c]\generator[a]+\generator[b]\generator[a]\generator[c]+\generator[c]\generator[a]\generator[b]+\generator[c]\generator[b]\generator[a]\bigr)\\
        &-\frac{1}{15}(3\sothreecasimirelement+1)\bigl(\delta_{ab}\generator[c]+\delta_{ac}\generator[b]+\delta_{bc}\generator[a]\bigr)
    \end{aligned}
\end{tblr}
\end{math}

    }
    \caption{Images of the multipoles $k=0,\dots,4$ on $k$-adic tensors.}\label{tab:multipole-images}
\end{center}
\end{table}

The multipoles are notable for several reasons. Firstly, they enjoy a simple relationship with $\casaction$, and possess some useful properties,
\begin{subequations}
\begin{gather}
        \sothreead[\sothreecasimirelement](\multipoletensor{a_{1}a_{2}...a_{j}})=-j(j+1)\multipoletensor{a_{1}a_{2}...a_{j}}\\[0.25\baselineskip]
    \multipoletensor{a_{\sigma(1)}a_{\sigma(2)}...a_{\sigma(j)}} = \multipoletensor{a_{1}a_{2}...a_{j}}\\
    \sothreesum*{a_{m},a_{n}}\delta_{a_{m}a_{n}}\multipoletensor{...a_{m}...a_{n}...}=0,\,\forall m\neq n.
\end{gather}
\end{subequations}
Taken together, these properties mean there are $2k+1$ linearly independent multipoles of order $k$.

Secondly, the structure of $\usutwo$ is sufficient to allow us to determine their associative products with respect to one another, which may be seen in \Cref{tab:multipole-multiplication}. It should be noted that this structure is determinable at the level of $\usutwo$, and so will be consistent across all derived algebras up to some details we will now see.

\begin{table}[!t]
\begin{center}
    \makebox[0cm]{

\begin{math}
    \begin{tblr}{colspec={c | c | c | c}, rowsep=0.5\baselineskip, hline{2-Y}={solid}, hline{2}={1pt}, vline{2}={1pt} }
        RC & \multipoletensor{} & \multipoletensor{e} & \multipoletensor{ef} \\
        \multipoletensor{} & \multipoletensor{} & \multipoletensor{e} & \multipoletensor{ef} \\
        \multipoletensor{a} & \multipoletensor{a} & \dipoledipole{a}{e}{s} & \dipolequadrupole{a}{e}{f}{s} \\
        \multipoletensor{ab} & \multipoletensor{ab} & \quadrupoledipole{a}{b}{e}{s} & \quadrupolequadrupole{a}{b}{e}{f}{s}
    \end{tblr}
\end{math}

    }
    \caption{Section of the multipole multiplication table}\label{tab:multipole-multiplication}
\end{center}
\end{table}

Finally, they allow us to resolve an arbitrary element of $\usutwo$ into an $\reals[\sothreecasimirelement]$-linear combination of multipoles, with $\reals[\sothreecasimirelement]$ the polynomial ring of $\sothreecasimirelement$ with real coefficients. This allows us to realise real algebraic descriptions of the spin degrees of freedom of non-relativistic systems.

\subsection{Spin Algebras}

Since we may decompose an arbitrary element of $\usutwo$ in terms of $\reals[\sothreecasimirelement]$ combinations of multipoles, the spaces of multipoles trivially intersect, and each space is closed under commutators with elements of $\sutwo$, we may perform quotients in which we set different families of multipoles to zero,
\begin{equation}\label{eqn:spin-algebra-definition}
    \spinalgebra{s}\cong\frac{\usutwo}{\ideal{\image{\multipole{2s+1}}}}.
\end{equation}

In this way, we define an infinite family of algebras in which only finitely many multipoles are non-zero. The multiplication between multipoles in each of these algebras is still given by \Cref{tab:multipole-multiplication}, with some multipoles set to zero as required. Remarkably, the quotient \eqref{eqn:spin-algebra-definition} also entails that the Casimir element $\sothreecasimirelement$ takes on a real scalar value: in $\spinalgebra{s}$, we have,
\begin{equation}
    \sothreecasimirelement=-s(s+1).
\end{equation}
Therefore, each $\spinalgebra{s}$ is in fact a finite-dimensional real algebra, as we hoped. A schematic of the content of the spin algebras and their unique identities is given in \Cref{tab:spin-algebras}.

\begin{table}[!t]
\begin{center}
    \makebox[0cm]{
        
\begin{math}
    \begin{tblr}{colspec={c | c c c},rowsep=0.5\baselineskip,hline{2}}
    \textup{Spin Algebra} & \textup{Content} & \textup{Defining Identity} & \textup{Consequence} \\
    \spinalgebra{0} & \sothreespinzeromultipoles & \multipoletensor{a}=0 & \sothreecasimirelement=0 \\
    \spinalgebra{\half} & \sothreespinhalfmultipoles{a} & \multipoletensor{ab}=0 & \sothreecasimirelement=-\frac{3}{4} \\
    \spinalgebra{1} & \sothreespinonemultipoles{a}{b} & \multipoletensor{abc}=0 & \sothreecasimirelement=-2 \\
    \vdots & \vdots & \vdots & \vdots \\
    \spinalgebra{s} & \set{\multipoletensor{},\multipoletensor{a_{1}},...,\multipoletensor{a_{1}...a_{2s}}} & \multipoletensor{a_{1}\dots a_{2s+1}}=0 & \sothreecasimirelement=-s(s+1)
    \end{tblr}
\end{math}

    }
    \caption{Schematic of the spin algebras with their unique identities}\label{tab:spin-algebras}
\end{center}
\end{table}

It is important to comment on the use of multipoles as a description for a system with non-relativistic spin within the real spin algebras $\spinalgebra{s}$. Firstly, with respect to the usual Hilbert space inner product on such systems, all odd-order multipoles are naturally anti-Hermitian, and all even-order naturally Hermitian. Since we associate observables in quantum mechanics with Hermitian operators, it is clear that on its own the $\spinalgebra{s}$ are not sufficient to describe spin as observed in the lab. Indeed, the condition that $\multipoletensor{ab}=0$ for spin-$\half$ systems entails that,
\begin{equation}
    0=\sothreegen[a]^{2}+\frac{1}{4};
\end{equation}
this means that the spectrum of $\sothreegen[a]$ is predicted ($\pm\unitimaginary\half$), but since we have no complex scalars in $\spinalgebra{\half}$ we have no access to the usual dipole eigenstates. This reveals a subtle quality of spin: its phenomenological character as angular momentum with eigenstates is a \textit{prediction} of quantum mechanics, not an assumption. This has been made manifest by our real algebraic approach, which must be combined with structure entailed by other structure within a more complete quantum theory to realise the phenomenon fully.

Secondly, since each algebra is finite-dimensional, we may use the multipole description as an operator basis for the endomorphisms which act on a spin-$s$ spinor. Here there is a subtlety: each order of multipole has $2k+1$ linearly independent objects, but there are $\frac{k(k+1)}{2}$ multipoles of the form $\set{\multipoletensor{a_{1}a_{2}\dots a_{k}}}$. This means what we might consider our ``basic'' multipoles, for example $\set{\multipoletensor{xx},\multipoletensor{xy},\multipoletensor{xz},\multipoletensor{yy},\multipoletensor{yz},\multipoletensor{zz}}$ for $k=2$, span their respective spaces but are not themselves a basis for it. There is also no canonical choice of $2k+1$ independent linear combinations of these that we can make. Contrasting this with the spherical tensors is instructive. In that case, there were a canonical (up to choice of scaling) set of $2k+1$ objects once the $k=1$ system was specified; however, the choice of $k=1$ system necessarily singled out a dipole element as a kind of zenith in space, breaking the symmetry of the problem. While any dipole element could be chosen, the formalism still required a choice in order to be defined. In contrast, the multipole description allows, but \textit{does not require}, a choice of zenith, and the lack of linear independence between the $\set{\multipoletensor{a_{1}a_{2}\dots a_{k}}}$ reflects the fact that the symmetry of our description remains unbroken. This offers great flexibility and descriptive power, as we may refine our description in myriad ways as a particular problem requires (the spherical tensor system also being accessible to a complexified system). It is worth noting that though we have worked with a particular basis of dipoles so far, each order of multipoles is accessible independently of this choice by constructing projection operators using $\casaction$, since each has a different eigenvalue with respect to this.

Finally, and perhaps most abstractly, the derivation of the spin algebras outlined here offers a way to access the irreducible representations of $\sutwo$ in a more direct way. Indeed, we required no eigenstates of our generators to exist, nor required states of any kind! The objects of each $\spinalgebra{s}$ have been fully accounted for in a manner consistent across different systems. The algebraic structure of $\spinalgebra{s}$ is structurally identical for each $s$ (\Cref{tab:multipole-multiplication}) with just the multipoles which are non-zero and the value of $\sothreecasimirelement$ in it varying. They have also offered an alternative characterisation of these systems to the eigenstate picture more directly in terms of their physically distinguishable observables: a spin-$s$ system has largest multipole observable of order $2s$, with the $2s+1$ and higher observables identically zero. The central role the multipoles play in this method and in defining the spin of a particular system motivate the study of these higher-order moments in experiment. The real algebraic approach underlying the derivation of the spin algebras is general, and can be applied to the study of other Lie groups; this reveals the possibility of avoiding the need to use existing representation theoretic results in the derivation of harmonic oscillator and hydrogen spectra as we previously required in \Cref{sec:harmonic-oscillator,sec:hydrogen}.

\section{Discussion and Conclusion}\label{sec:discussion-conclusion}

In the preceding sections, we have presented a window into a much larger world of algebraic approaches to quantum mechanics. To comprehensively review this area of research would require much more space than available to us, but for the interested reader we provided a sample of active topics to explore further. 

When someone discusses algebraic approaches to quantum theory, commonly they are referring to ``Algebraic Quantum Field Theory''~\cite{fewster_algebraic_2020}. This considers algebras of bounded linear operators defined over open sets of a spacetime manifold, and aims to provide both a rigorous foundation for quantum field theory and expand our current knowledge in that domain. 

Another popular area enjoying significant interest in both mathematics and physics is ``Deformation Quantisation''~\cite{deng_finite_2008}. While yielding novel mathematical connections of interest in pure maths, this theory also generalises what may be considered the transition from classical to quantum: the deforming of the position and momentum commutation relations to be non-commutative by introducing a new parameter $\hbar$. Of particular interest to physics, deformation quantisation provides a programme for ``quantising'' structures familiar to physical models. For example, if one deforms the commutative algebra of phase space functions, we arrive at the theory of ``star products'' and the ``Moyal bracket''~\cite{zachos_survey_2001, hiley_relationship_2015, de_gosson_principles_2016}. This theory is equivalent to traditional quantum mechanics but formulated on classical phase space instead of the usual configuration space.

Another example that has been receiving much attention due to its geometric language are ``Projective and Conformal Geometric Algebra''. These theories, based on Clifford algebras (which are deformations of the exterior algebra), have enjoyed success in geometric and mechanical applications~\cite{hitzer_survey_2024, dileri_projective_2026, navrat_hamiltonian_2026}. A related but distinct approach has been taken by Furey, who uses the normed division algebras to model unifications of the standard model of particle physics~\cite{furey_algebraic_2025, furey_algebraic_2025-1, furey_algebraic_2025-2}.

Beyond these theories are more exploratory approaches such as supersymmetric quantum mechanics~\cite{wipf_algebraic_2006, levai_solvable_1994}, and the vertex algebras of two-dimensional conformal field theory~\cite{kac_vertex_1998} and string theory. Perhaps most speculative of all, we may follow Okubo's lead~\cite{okubo_introduction_1995} and consider further relaxing an assumption of quantum mechanics currently held sacred. Just as classical mechanics was superseded by quantum through the relaxation of the commutative law of numbers~\cite{dirac_fundamental_1925, dirac_quantum_1926-1}, maybe we too could move beyond quantum mechanics by either relaxing associativity to alternativity, so considering Malcev algebras and Moufang loops instead of Lie algebras and Lie groups; or by abandoning the binary product between operators in favour of ternary or greater products, as in Nambu mechanics~\cite{curtright_classical_2003}.

In this paper, we presented a collection of well-studied quantum systems: the harmonic oscillator, hydrogen atom, optical components, and particle spin. Each was analysed by algebraic means in a number of ways highlighting the algebraic relationships between operators as responsible for encoding the physical character of each system; most systems were directly solvable once this structure was identified. The Lie algebra $\sotwoone\cong\suoneone$ appeared across multiple systems in its role as ``spectrum generating algebra'', and hidden $\specialorthogonalgroup{4}$ and $\specialunitarygroup{3}$ symmetries were displayed in the hydrogenic atom and isotropic harmonic oscillator respectively. We exemplified the ``Hong-Ou-Mandel'' effect as a tangible physical consequence of the algebraic relationship between creation and annihilation operators in quantum optics, and compared a more elementary algebraic perspective on the nature of spin in quantum mechanics with other algebraic approaches. We finally discussed areas of active research within the algebraic quantum community.

\section*{Acknowledgements}

The authors acknowledge support from a Royal Society Research Grant (RG/R1/251590), an EPSRC Mathematical Sciences Small Grant (UKRI3647), and from JRH's EPSRC Quantum Technologies Career Acceleration Fellowship (UKRI1217). PTJB thanks Hamish Todd for stimulating discussion, and Maja Hajtanian for her support while writing this paper.

{%
\raggedright%
\printbibliography%
}%

\end{document}